\documentclass[aps,amssymb,amsmath,prl,superscriptaddress,reprint,noshowpacs]{revtex4-1}

\usepackage{times}
\usepackage{amssymb,amsmath,graphicx}
\usepackage[usenames]{color}
\usepackage{bbold}
\usepackage{siunitx}
\usepackage{braket}
\usepackage{comment}
\usepackage[T1]{fontenc}
\usepackage[utf8]{inputenc}
\usepackage[english]{babel}
\usepackage{etoolbox}
\usepackage{lipsum} 

\usepackage{hyperref}
\hypersetup{colorlinks=true, citecolor=blue, urlcolor=blue, linkcolor=blue}

\usepackage[normalem]{ulem}
\usepackage{lineno}

\newcommand{\Om}{\Omega_\mathrm{m}}
\newcommand{\Oe}{\Omega_\mathrm{e}}

\newcommand{\Gtot}{\Gamma_\mathrm{tot}}

\newcommand{\Gqba}{\Gamma_\mathrm{qba}}

\newcommand{\zzpf}{z_\mathrm{zpf}}
\newcommand{\pzpf}{p_\mathrm{zpf}}

\usepackage{textcomp}

\newcommand{\commentOut}[1]{}
\usepackage{scalefnt}   

\newcommand{\affil}{Photonics Laboratory, ETH Zürich, CH-8093 Zürich, Switzerland}
\newcommand{\affilQC}{Quantum Center, ETH Zürich, CH-8093 Zürich, Switzerland}
\newcommand{\affilICFO}{
ICFO -- Institut de Ciencies Fotoniques, The Barcelona Institute of Science and Technology, Castelldefels,
Barcelona 08860, Spain}
\newcommand{\affilICREA}{
ICREA -- Institucio Catalana de Recerca i Estudis Avancats, Barcelona 08010, Spain}
\newcommand{\affilIQOQI}{
Institute for Quantum Optics and Quantum Information of the Austrian Academy of Sciences, A-6020 Innsbruck, Austria}

\begin{document}
\scalefont{1.05}
\title{Quantum Delocalization of a Levitated Nanoparticle}

\author{M. Rossi}
\altaffiliation{Present address: Kavli Institute of Nanoscience, Department of Quantum Nanoscience, Delft University of Technology, 2628CJ Delft, The Netherlands}
\affiliation{\affil}
\affiliation{\affilQC}

\author{A. Militaru}
\altaffiliation{Present address: Institute of Science and Technology Austria, Am Campus 1, 3400 Klosterneuburg, Austria}
\affiliation{\affil}
\affiliation{\affilQC}

\author{N. Carlon Zambon}
\affiliation{\affil}
\affiliation{\affilQC}

\author{\\A.~Riera-Campeny}
\affiliation{\affilICFO}
\affiliation{\affilIQOQI}

\author{O. Romero-Isart}
\affiliation{\affilICFO}
\affiliation{\affilICREA}

\author{M. Frimmer}
\affiliation{\affil}
\affiliation{\affilQC}

\author{L. Novotny}
\affiliation{\affil}
\affiliation{\affilQC}

\begin{abstract}
Every massive particle behaves like a wave, according to quantum physics. Yet, this characteristic wave nature has only been observed in double-slit experiments with microscopic systems, such as atoms and molecules \cite{keith_interferometer_1991, fein_quantum_2019}. The key aspect is that the wavefunction describing the motion of these systems extends coherently over a distance comparable to the  slit separation, much larger than the size of the system itself \cite{nairz_quantum_2003}.
Preparing these states of more massive and complex objects remains an outstanding challenge.
While the motion of solid-state oscillators can now be controlled at the level of single quanta, their coherence length remains comparable to the zero-point motion, limited to subatomic distances \cite{oconnell_quantum_2010, teufel_sideband_2011, rossi_measurement-based_2018, bild_schrodinger_2023}. 
Here, we prepare a delocalized state of a levitating solid-state nanosphere with coherence length exceeding the zero-point motion. We first cool its motion to the ground state. Then, by modulating the stiffness of the confinement potential, we achieve more than a threefold increment of the initial coherence length with minimal added noise. Optical levitation gives us the necessary control over the confinement that other mechanical platforms lack. Our work is a stepping stone towards the generation of delocalization scales comparable to the object size, a crucial regime for macroscopic quantum experiments \cite{romero-isart_large_2011}, and towards quantum-enhanced force sensing with levitated particles~\cite{burd_quantum_2019}. 
\end{abstract}

\maketitle
Quantum mechanics brought a dose of fuzziness to the description of the physical world. 
This vagueness is fundamental and not due to our lack of knowledge \cite{heisenberg_uber_1927}. 
The paradigmatic example is an electron bound to an atomic nucleus. Such an electron cannot be pinpointed in space but should be understood as delocalized over its orbit.
\\
The delocalization of objects of increasing size is of twofold interest.
On the one hand, there is a fundamental interest in testing quantum mechanics on ever larger systems \cite{ghirardi_unified_1986, romero-isart_quantum_2011, bassi_models_2013, arndt_testing_2014}.
On the other hand, these systems feature enhanced sensing capabilities, which find use in gravitational measurements \cite{westphal_measurement_2021, weiss_large_2021, cosco_enhanced_2021} and search of novel physics \cite{moore_searching_2021}.
For delocalization to be useful, one needs to achieve it in a coherent manner.
That is, it must come from the fundamental quantum uncertainty encoded in the wavefunction, a so-called pure state, and not from added noise due to experimental imperfections \cite{weiss_large_2021}.
The coherence length appropriately quantifies this quantum delocalization \cite{joos_emergence_1985, nairz_quantum_2003, schlosshauer_decoherence_2007, romero-isart_quantum_2011}.
Increasing the coherence length to large distances is a daunting task because of decoherence \cite{schlosshauer_decoherence_2007}.
For instance, a single scattering event, e.g., from a surrounding gas molecule, can collapse the motional state of a macroscopic particle into a well localized region \cite{uys_matter-wave_2005-1}.
\\
Delocalized states are routinely prepared in atomic physics \cite{meekhof_generation_1996, xin_rapid_2021} and matter-wave interference experiments with neutrons \cite{rauch_test_1974}, atoms \cite{keith_interferometer_1991}, and molecules \cite{fein_quantum_2019}. 
Achieving a similar level of control for more macroscopic objects remains an outstanding challenge due to their initially low coherence length and strong interaction with the surrounding environment \cite{aspelmeyer_cavity_2014,gonzalez-ballestero_levitodynamics_2021}. 
A first important step in this direction is ground-state cooling of mechanical systems, see for instance Refs.~\cite{chan_laser_2011, teufel_sideband_2011, rossi_measurement-based_2018, delic_cooling_2020, magrini_real-time_2021, tebbenjohanns_quantum_2021}.
For massive objects, however, the quantum delocalization achievable even in an ideal zero temperature bath is limited by the zero-point motion $\zzpf=\sqrt{\hbar/(m\Om)}$, where $
\hbar$, $m$ and $\Om$ are, respectively, the reduced Planck constant and the oscillator mass and resonance frequency \cite{oconnell_quantum_2010, teufel_sideband_2011, rossi_measurement-based_2018, bild_schrodinger_2023}.
Here, we coherently delocalize the state of a levitated nanomechanical oscillator beyond the limit imposed by the zero-point motion.
We achieve this delocalization by controlling the confinement potential.
We verify the state prepared by combining  a quantum-limited position measurement with a retrodiction-based estimation.

The coherence length describes the distance for which spatial correlation persists.
For a given state described by the density matrix $\rho$, the correlation function at point $z$ and $z'$ is $\langle z|\rho|z'\rangle$.
The relevant states for current experiments are the so-called Gaussian states, which feature an exponential decay $\exp[-(z-z')^2/\xi^2]$ in their correlation function.
We refer to the parameter $\xi=\sqrt{8}P\Delta z$ as the coherence length \cite{romero-isart_quantum_2011}, where $P=\mathrm{Tr}(\rho^2)$ is the state purity and $\Delta z$ is the position standard deviation, termed also uncertainty.
For a resonator in its ground state  $\xi=2\zzpf$.

To increase the coherence length beyond the zero-point motion, we need to increase the position uncertainty while retaining a high purity.
A viable approach to enlarge this uncertainty is to change the resonance frequency, e.g., via a periodic modulation \cite{rugar_mechanical_1991,blencowe_quantum_2000, mari_gently_2009, asjad_robust_2014}.
The frequency of tethered nanomechanical oscillators, however, can be tuned only over a limited range \cite{vinante_feedback-enhanced_2013, poot_deep_2015, mashaal_strong_2024}.
This limits the delocalization achievable within the coherence time of the oscillator.
We employ a different class of mechanical resonators: we levitate a nanosphere in an optical tweezer.
Its center of mass behaves as a harmonic oscillator, with an optically defined spring that can be arbitrarily tuned and even switched off \cite{romero-isart_large_2011,hebestreit_sensing_2018, frimmer_rapid_2019}. 
\\
We start by cooling the nanosphere’s motion to the ground state of the tweezer potential. 
By suddenly lowering the laser power, the restoring force decreases and hence the nanosphere delocalizes in space.
This also reduces recoil heating, which is the main source of decoherence in ultra-high vacuum.
During this step,  the dynamics resembles the expansion phase in a matter-wave interferometer \cite{nairz_quantum_2003}.
Toggling the tweezer back to its original power at the moment of maximum expansion allows us to recapture the delocalized nanosphere.
The harmonic evolution, then, induces alternation between states highly delocalized but with precise momentum, and vice versa. 
Eventually, photon scattering off the particle destroys the coherent delocalization \cite{romero-isart_quantum_2011}.
By collecting these scattered photons, we measure the nanoparticle's center of mass motion and its maximum expansion.

\begin{figure}[tbp]
\centering
\includegraphics[width=86mm]{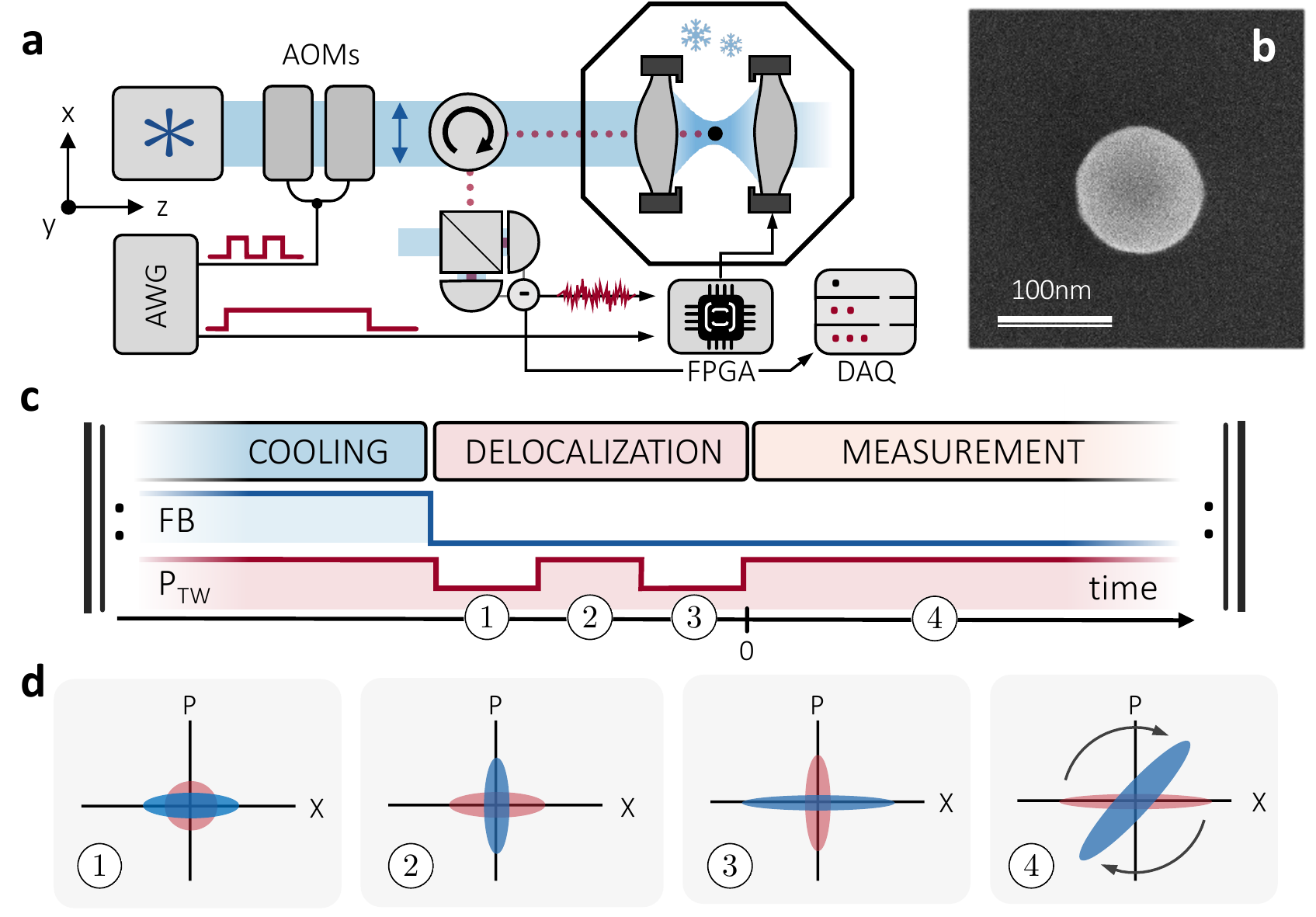}
\caption{\textbf{Experimental setup and expansion protocol.} 
\textbf{a} Schematic overview of the optical and electronic setup. 
The power of the laser forming the tweezer is modulated by square waves.
The nanoparticle's motion is monitored by detecting the scattered photons and the corresponding signal is processed and acquired by digital electronics.
AOM, acousto-optic modulator; AWG, arbitrary waveform generator; DAQ, data acquisition card; FPGA, field-programmable gate array.
\textbf{b} Scanning electron microscope image of a nanoparticle.
\textbf{c} Steps for delocalizing the nanoparticle.
Feedback cooling is switched off during "delocalization" and "measurement".
The tweezer is modulated during "delocalization" only.
FB, feedback; $\mathrm{P_{TW}}$ tweezer power.
\textbf{d} Phase space distribution of the mechanical state during each step. 
The red (blue) ellipses represent the covariance matrix of the Gaussian state at the beginning (end) of the numbered phase of the experiment, in correspondence with \textbf{c}.}
\label{fig:experimental_setup}
\end{figure}
The essential elements of the experiment are illustrated in Fig.~\ref{fig:experimental_setup}a. 
We levitate a silica nanoparticle (nominal diameter \SI{100}{nm}, mass $m \approx \SI{1.2}{fg}$) inside a vacuum chamber by means of an optical tweezer propagating along the $z$ axis.
Figure~\ref{fig:experimental_setup}b shows a scanning electron microscope image of a typical particle. 
The optical tweezer is formed by focusing a linearly $x$-polarized laser beam (wavelength \SI{1550}{nm}, power $0.66\pm0.01~\mathrm{W}$) to a diffraction-limited volume with a single aspheric lens (numerical aperture 0.8).
For displacements of the nanoparticle much smaller than the laser wavelength, the center of mass motion along the three axes is decoupled from each other and harmonic with resonance frequencies $2\pi\times\{56.5,\,151.0,\,186.5\}\,$kHz, for the axes $\{z, x, y\}$ respectively.
In this work, we focus solely on the longitudinal motion along the $z$ axis and refer to its resonance frequency as $\Om$.
The lens holder is attached to the cold plate of a closed-cycle cryostat, which provides a temperature around the particle of \SI{7}{K} and a pressure below $10^{-9}$\,mbar.
Operating the system in ultra-high-vacuum makes the decoherence due to collisions with the background gas negligible (see Supplementary Material) \cite{tebbenjohanns_quantum_2021}.
\\
We monitor the displacement of the nanoparticle by detecting interferometrically the phase shift imprinted on backscattered photons.
This detection scheme is the most sensitive to measure the longitudinal motion \cite{tebbenjohanns_optimal_2019}.
From our position measurements, we estimate a total decoherence rate of $\Gamma_\mathrm{tot}=(23.7\pm 1.6)\times10^3\,\mathrm{phonons\,s^{-1}}$ and use this parameter to calibrate the data presented hereafter (see Supplementary Material).
The decoherence in our experiment is dominated by the random recoil due to photon scattering, also known as photon recoil (rate $\Gqba$) \cite{tebbenjohanns_quantum_2021,magrini_real-time_2021}.
\\
To cool the three motional modes of the nanoparticle and lower their energy, we generate a feedback signal from the acquired position measurement.
By driving three pairs of orthogonal electrodes and exploiting the electric charge carried by the nanoparticle, we can cool the transverse modes to a low occupancy thermal state, and the longitudinal one close to its ground state, with a residual average occupancy of $\overline{n} = 0.68\pm0.09$, measured via standard heterodyne thermometry techniques \cite{purdy_quantum_2017, tebbenjohanns_motional_2020}.
In the Supplementary, we also show additional datasets starting from different initial occupancies for the longitudinal motion.
\\
After cooling, the longitudinal motion of the nanoparticle is well localized. The uncertainty in its position is $\Delta z = \sqrt{(\overline{n}+1/2)} \zzpf \approx \SI{17}{pm}$ and the initial coherence length is $\xi_0\approx21\,\mathrm{pm}$, since the purity for a thermal state is $P=(2\bar{n}+1)^{-1}$.
In order to explore quantum mechanics at larger scales, we need to increase $\xi$, in other words to delocalize the nanoparticle motion while retaining its purity \cite{romero-isart_coherent_2017,weiss_large_2021}.
We do so by suddenly softening the trapping potential by reducing the focal intensity of the optical tweezer.
We perform this by means of an intensity modulator placed in the path of the optical tweezer before reaching the nanoparticle (see Supplementary Material).
The switching time of this modulator is around \SI{100}{ns}, a much faster timescale than the nanoparticle oscillation period of $\SI{18}{\mu s}$.

We design our experiment according to the following three-step protocol (shown in Fig.~\ref{fig:experimental_setup}c).
First, in the "cooling" step, we tune the laser intensity to the maximum value, corresponding to a mechanical resonance frequency $\Om$, and feedback cool the motion close to its ground state, with residual occupancy $\bar{n}$.
Once the steady state is attained, we move to the second step, the "delocalization". Here we switch the feedback off and modulate the trapping potential with a pulse sequence.
In the last step, "measurement", we reconstruct the delocalized state.
To that end, we toggle the laser intensity back to its maximum value and acquire the homodyne signal without applying any feedback.
We use the data acquired in this step to estimate the state generated at the end of the pulse sequence.
\\
To delocalize the nanoparticle, we employ a 2-pulse sequence.
These pulses (labels 1 and 3 in Fig.~\ref{fig:experimental_setup}c) lower the mechanical resonance frequency to $\Oe = \Om/r$, where $r$ denotes the frequency ratio. 
Their duration is $\pi/(2\Oe)$, i.e., a quarter of the lower mechanical period $2\pi/\Oe$.
During both pulses, the position uncertainty increases proportionally to the frequency ratio $r$, as sketched in Fig.~\ref{fig:experimental_setup}d \cite{janszky_strong_1992}.
Due to the lower impinging power, the decoherence rate due to photon recoil is reduced by a factor $r$.
We space these two pulses by a quarter of the original period, i.e., $\pi/(2\Om)$ (label 2 in Fig.~\ref{fig:experimental_setup}c).
During the time between the pulses, the nanoparticle oscillates at a rate $\Om$.
We introduce this spacing between the pulses to flip the sign of the average momentum that the nanoparticle has gained during the first pulse, e.g., due to stray static electric fields.
In this way, during the last pulse the nanoparticle continues to delocalize while containing the detrimental effect that drifting stray forces may have on the final state (see Supplementary Material).

\begin{figure}[tbp]
\begin{center}
\includegraphics[width=86mm]{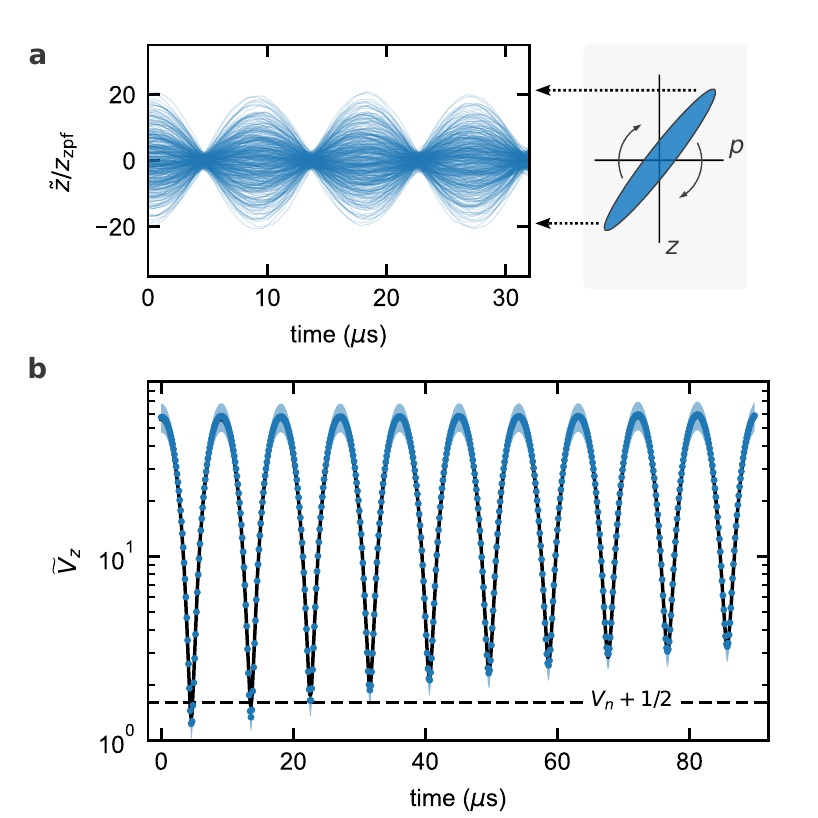}
\caption{\textbf{State evolution after recapture.}
\textbf{a} Ensemble of about 400 filtered trajectories.
The zero of the time axis corresponds to the beginning of the "measurement" step.
The phase space diagram on the right illustrates the dynamics of the generated state.
\textbf{b} Ensemble variance of the position as a function of time (circles), and fit (black line). 
The blue shaded area represents the 95\% confidence on the data.
The horizontal dashed line represents the sum between the variance of the zero point motion ($1/2$) and our measurement uncertainty ($V_n$).}
\label{fig:expanded_variance}
\end{center}
\end{figure}

\begin{figure}[tbp]
\begin{center}
\includegraphics[width=86mm]{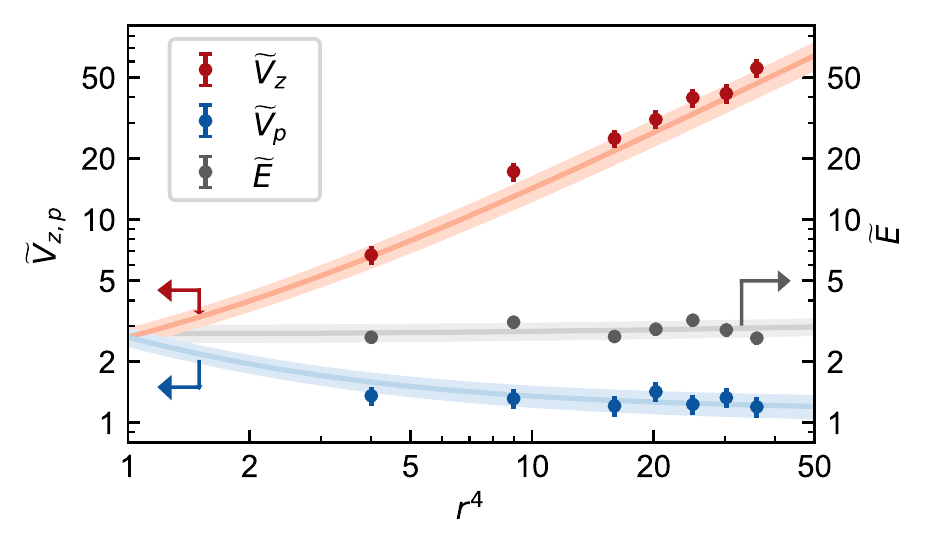}
\caption{\textbf{Estimated variance and energy after delocalization and recompression.}
The red and blue circles are, respectively, the estimated position and momentum variances after the delocalization, and refer to the left vertical axis. 
The datapoints for $\widetilde{V}_z$ and $\widetilde{V}_p$ at the highest $r$ are obtained from the data shown in Fig.~\ref{fig:expanded_variance}.
The gray circles represent the total energy of the particle after recompression and refer to the right vertical axis. 
The error bars on the experimental data represent the 95\% confidence interval deduced from the fit parameters. 
The solid coloured lines are theoretical predictions and the shaded areas reflect the 95\% confidence interval of the model parameters.}
\label{fig:expansion_vs_r}
\end{center}
\end{figure}

We choose $r$=2.45 and repeat this experiment about 400 times.
In Fig.~\ref{fig:expanded_variance}a we show, for each repetition, the inferred nanoparticle's motion in units of zero point motion.
The initial time $t=0\,\mathrm{\mu s}$ corresponds to the beginning of the "measurement" step.
To extract the trajectory, we filter in two steps the raw homodyne signal acquired during the "measurement" step.
First, we apply backward in time a pair of second-order high-pass filters to reject noise in the low frequency part of the spectrum (cutoff $3\,\mathrm{kHz}$ and $9\,\mathrm{kHz}$).
Then, we apply a retrodiction filter based on a stochastic master equation to optimally infer the motion down to the beginning of the "measurement" step \cite{rossi_observing_2019, lammers_quantum_2024}.
Each of the trajectories shown in Fig.~\ref{fig:expanded_variance}a is the output of such a retrodiction filter.
\\
From the data, we can qualitatively deduce that the position uncertainty (the spread of trajectories) oscillates at $2\Om$.
This is consistent with an asymmetric position-momentum (phase space) distribution.
This distribution rotates in the phase space at a rate $\Om$.
Consequently, its position marginal oscillates at twice that rate (cf. inset of Fig.~\ref{fig:expanded_variance}a).
We calculate at each time step the position ensemble variance, $\widetilde{V}_z$, over the 400 realizations shown in Fig.~\ref{fig:expanded_variance}b.
Hereafter, we refer to estimated quantities (either from experiment or theory) with a tilde.
In addition to the oscillations at $2\Om$, we observe that the minimum variance increases with time.
This increase is due to the photon recoil.
To quantify the maximum delocalization, we fit the data up to $360\,\mathrm{\mu s}$ to the following model (see Supplementary Material for its derivation)
\begin{eqnarray}
\label{eq: variance in time}
\widetilde{V}_z(t) &= \widetilde{V}_z(0)\cos\left(\Om t\right)^2+\widetilde{V}_p(0)\sin\left(\Om t\right)^2 + \\
&+\widetilde{C}_{zp}(0)\sin\left(2\Om t\right) + \Gqba \left(t-\frac{\sin\left(2\Om t\right)}{2\Om}\right), \nonumber
\end{eqnarray}
where $\widetilde{V}_z(0), \widetilde{V}_p(0)$ and $\widetilde{C}_{zp}(0)$ are, respectively, the estimated position and momentum variances and the symmetrized covariance at time $t=0\,\mathrm{\mu s}$.
In Eq.~\eqref{eq: variance in time}, the position and momentum are normalized to their corresponding zero point motion $\zzpf$ and $\pzpf=\hbar\zzpf^{-1}$, respectively.
We notice that both variance estimates are biased by the same estimation noise $V_n$, i.e., $\widetilde{V}_z(0)=V_z(0)+V_{n}$ and $\widetilde{V}_p(0)=V_p(0)+V_{n}$.
This bias originates from imprecision noise in the measurement that the filter cannot reject \cite{wieczorek_optimal_2015, rossi_observing_2019, lammers_quantum_2024}.
From our parameters, we find that $V_n=(1.1\pm0.1)$.
For reference a perfectly efficient measurement would yield $V_n=0.5$.
This is because our apparatus operates in the weak measurement regime ($\Gqba/\Om\approx0.07\ll1$), and is effectively equivalent to a simultaneous position and momentum measurement \cite{doherty_quantum_2012}.
The finite estimation noise ensures that the Heisenberg principle is fulfilled.
From the fits, we obtain the covariance matrix of the delocalized mechanical state.
In principle, this covariance matrix should be diagonal, with the position variance as the largest value.
In practice, small inaccuracies in the pulses' duration might generate a mechanical state with finite correlations between position and momentum, with the largest delocalization in a rotated quadrature ($\delta\tilde{\theta}\approx0.02 \pi$).
We find this delocalization by extracting the maximum eigenvalues of the estimated covariance matrix. 
For the sake of simplicity, we keep referring to the maximum and minimum eigenvalues as the position and momentum variance, respectively.
For this experiment, we find $\widetilde{V}_z=56\pm6$. This corresponds to a position uncertainty of $\Delta z= (120\pm6)\,\mathrm{pm}$, seven times the initial uncertainty of $17\,\mathrm{pm}$.
\\
We repeat the measurement outlined above for different frequency ratios $r$ and show the results in Fig.~\ref{fig:expansion_vs_r}.
The red and blue circles represent, respectively, the maximum and minimum eigenvalues of the estimated covariance matrix.
We observe that the position variance scales with the fourth power of the frequency ratio, $r^4$, whereas the estimated momentum variance approaches the estimation noise $V_n$.
To account for these data, we model our experiment by standard optomechanical theory in free space \cite{militaru_ponderomotive_2022}.
From this model, we predict
\begin{subequations}
\label{eq: state prediction}
\begin{align}
\widetilde{V}_{z,\mathrm{th}} &= V_n + r^4 V_0 + \frac{\Gqba}{\Om}\frac{\pi}{2}\left(r+r^2+r^3\right),\label{eq:vzest_th}\\
\widetilde{V}_{p,\mathrm{th}} &= V_n + r^{-4} V_0 + \frac{\Gqba}{\Om}\frac{\pi}{2}\left(r^{-1}+r^{-2}+r^{-3}\right),\label{eq:vpst_th}
\end{align}
\end{subequations}
where $V_0=\bar{n}+1/2$ and $\bar{n}$ is the initial phonon occupancy (due to feedback cooling).
In particular, note that we assume photon recoil as the only significant decoherence throughout the experiment, i.e., $\Gtot\approx\Gqba$.
The solid lines in Fig.~\ref{fig:expansion_vs_r} are predictions from the model without fitting parameters.
The shaded area reflects the uncertainty in the independently measured parameters used to evaluate Eqs.~\eqref{eq: state prediction}.
The good agreement between experiment and theory supports the assumption that decoherence effects besides photon recoil are negligible in our experiment.
\\
These data show an increase in the position uncertainty. 
To verify that the prepared state has not lost coherence,  we perform the following experiment: we delocalize and immediately compress the nanoparticle state back \cite{weiss_large_2021}.
For this experiment, we change the duration of the pulse sequence: we double the length of the two pulses, i.e., $\pi/\Oe$, and shorten their spacing to the minimum possible amount in practice, i.e., $\sim200\,\mathrm{ns}$ limited by the AOM switching time.
In this way, we approach a single pulse of duration $2\pi/\Oe$.
Since this duration is the period of the dynamics during the pulse, the final state equals the initial one.
We leave the measurement and estimation steps unchanged.
In Fig.~\ref{fig:expansion_vs_r} we show the estimated energy of the final state as gray circles.
We derive the energy from the position and momentum variance as $\widetilde{E}=(\widetilde{V}_z+\widetilde{V}_p)/2$.
In absence of decoherence, the total mechanical energy at the end is unchanged and the purity unaffected.
In practice, the residual decoherence due to photon recoil slightly increases the nanoparticle energy, after the protocol.
From our model, we predict for the estimated energy after the protocol $\widetilde{E}_\mathrm{th}=V_n+V_0+\Gqba(\Om \pi)^{-1} \left(r+r^{-1}\right)$.
We compute this prediction and show it as a solid line in Fig.~\ref{fig:expansion_vs_r}, with the shaded area reflecting the uncertainty in the parameters used to evaluate it.
The good agreement between prediction and experiment supports, once more, our assumption that photon recoil is the dominant process during delocalization.

\begin{figure}[tbp]
\begin{center}
\includegraphics[width=86mm]{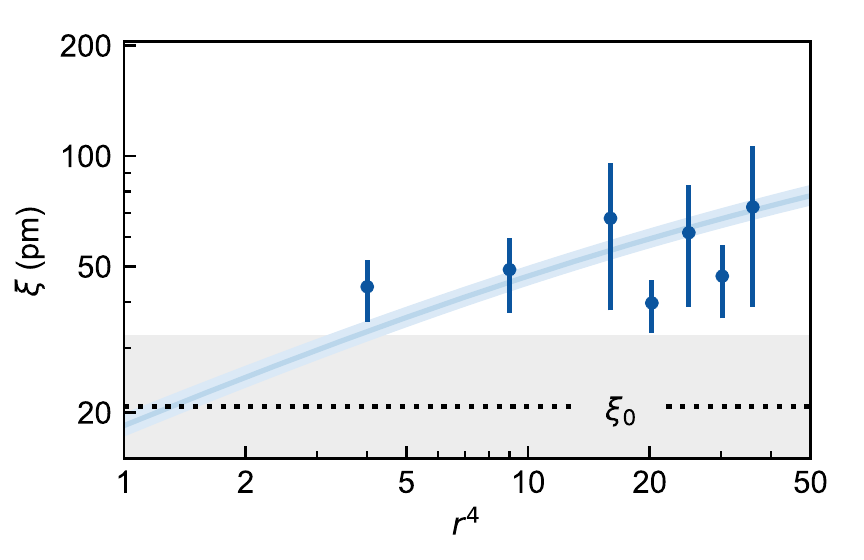}
\caption{\textbf{Inferred coherence length.}
The circles are inferred coherence lengths and the error bars represent the 95\% confidence interval of the estimates.
The blue solid line is the prediction from our model, and the shaded area reflects the 95\% confidence interval in the model parameters.
The gray area bounds the region of coherence lengths obtainable with ground-state cooling.
The dashed line marks the initial coherence length.}
\label{fig:coherence length}
\end{center}
\end{figure}

To further support our claim of having prepared a state delocalized beyond the zero-point motion, we infer from the previous dataset the coherence length.
The state we generate presents, at time $0\,\mathrm{\mu s}$, negligible correlations between position and momentum.
Thus, the coherence length can be further simplified and expressed as $\xi=\zzpf\sqrt{2/V_p}$.
To infer the  momentum variance, we subtract from the estimated one $\widetilde{V}_p$ (from Fig.~\ref{fig:expansion_vs_r})  the estimation noise $V_n$.
In order to get meaningful errorbars for the variance, i.e., corresponding to a physically realizable state, we use a positive semidefinite optimization to perform the subtraction and a Montecarlo method to obtain a 95\% confidence interval \cite{kotler_direct_2021} (see also Supplemental Material).
We show in 
Fig.~\ref{fig:coherence length} the inferred coherence length for the different ratios $r$.
We observe that the inferred coherence length is always larger than its initial value.
In particular, we notice that for large $r$ it exceeds the coherence length of the ground state.
For the highest $r$, we have $\xi=(73\pm34)\,\mathrm{pm}$, more than a threefold increase compared to the initial state ($\xi_0\approx~21\,\mathrm{pm}$) and doubling the zero-temperature state coherence length ($\xi_{|0\rangle}\approx~32\,\mathrm{pm}$).
Our system exceeds the ground state coherence length due to the squeezed momentum fluctuations below the zero point value \cite{wollman_quantum_2015, pirkkalainen_squeezing_2015, youssefi_squeezed_2023}. Indeed, we find that momentum variance is squeezed by $-7.2^{+2.9}_{-11.5}\,\mathrm{dB}$ below its zero-point value.

Our results show a coherent delocalization of the position of a levitated nanosphere to $73\,\mathrm{pm}$.
This stems from the increased position uncertainty and a preserved state purity.
The observed excess noise is consistent with decoherence caused by quantum backaction due to photon scattering during the delocalization protocol.
To go to even larger coherence length, the next step to reduce this decoherence by implementing "dark" traps for the expansion protocol \cite{pino_-chip_2018,roda-llordes_macroscopic_2023}, such as RF traps \cite{conangla_extending_2020, bykov_hybrid_2022}. 
The use of such a hybrid trap has recently been demonstrated in the context of  classical state expansion \cite{bonvin_state_2023}.
To the best of our knowledge, there are no fundamental limitations to the application of the quantum control techniques developed in this work with the hybrid trap of Ref.~\cite{bonvin_state_2023}. 
The squeezing that underlies the enhanced coherence length  provides an advantage in force sensing \cite{burd_quantum_2019}, e.g., for the detection of small electric forces. 
Moreover, the methods presented here are important for elucidating open questions at the interface of quantum physics and gravity, such as the gravitational coupling of two delocalized quantum masses \cite{dewitt_role_2011, marletto_gravitationally_2017, bose_spin_2017, belenchia_quantum_2018}.

\textbf{Acknowledgments} We thank the Q-Xtreme synergy group and our colleagues at the ETH photonics laboratory for fruitful discussions. This research has been supported by the European Research Council (ERC) under the grant agreement No. [951234] (Q-Xtreme ERC-2020-SyG), the Swiss SERI Quantum Initiative (grant no. UeM019-2) and the Swiss National Science Foundation (grant no. 51NF40-160591). N.C.Z. thanks for support through an ETH Fellowship (grant no. 222-1 FEL-30). A.R.-C. acknowledges funding from the European Union’s Horizon 2020 research and innovation programme under the Marie Skłodowska-Curie grant agreement No. [101103589] (DecoXtreme, HORIZON-MSCA-2022-PF-01-01).

\textbf{Author Contributions} 
M.R. led the project.
M.R., A.M. and N.C.Z. designed the experiment, built the setup, acquired and analyzed the data in equal contribution. Together with M.F. and L.N., they discussed the results. A.R.-C. and O.R.-I. provided support with the theoretical modeling. All authors contributed to writing the manuscript. O.R.-I. and L.N. gave the research direction and vision. L.N. supervised the project.

\bibliography{main.bib}
\bibliographystyle{naturemag}

\end{document}


\title{Supplemental Material:\\Quantum Delocalization of a Levitated Nanoparticle}
\author{M. Rossi}
\altaffiliation{Present address: Kavli Institute of Nanoscience, Department of Quantum Nanoscience, Delft University of Technology, 2628CJ Delft, The Netherlands}
\affiliation{\affil}
\affiliation{\affilQC}

\author{A. Militaru}
\altaffiliation{Present address: Institute of Science and Technology Austria, Am Campus 1, 3400 Klosterneuburg, Austria}
\affiliation{\affil}
\affiliation{\affilQC}

\author{N. Carlon Zambon}
\affiliation{\affil}
\affiliation{\affilQC}

\author{\\A.~Riera-Campeny}
\affiliation{\affilICFO}
\affiliation{\affilIQOQI}

\author{O. Romero-Isart}
\affiliation{\affilICFO}
\affiliation{\affilICREA}

\author{M. Frimmer}
\affiliation{\affil}
\affiliation{\affilQC}

\author{L. Novotny}
\affiliation{\affil}
\affiliation{\affilQC}
\maketitle

\tableofcontents
\clearpage
\newpage

\section{Theory}

In this section, we derive the main equations used to fit our data. We start from the quantum Langevin dynamics to which the levitated particle is subject, and compute the time evolution of the operators following a sudden change in the mechanical resonance frequency. For simplicity, we only consider the longitudinal motion $z$, and assume that the transversal axes remain always decoupled from it throughout the protocol considered. Moreover, we neglect any effect of the thermal bath associated with the background gas, in other words we assume that the decoherence of the system is only given by the photon recoil introduced by the tweezer laser. Both assumptions are well justified for the parameters used in our experiments.

\subsection{State evolution after potential softening}\label{sec:theory_eom}
The starting point is the quantum Langevin equation for a nanoparticle optically trapped in free space. More details and the derivation can be found in one of our previous works \cite{militaru_ponderomotive_2022}.
The delocalization beyond the initial state arises from the dynamics generated by making the confinement potential shallower, i.e., changing the mechanical resonance frequency from the initial value $\Om$ to the lower one $\Oe$, where $r=\Om/\Oe$ is the frequency ratio.
The equations of motion in the shallower potential are 
\begin{subequations}
    \label{eq: second EOM}
    \begin{align}
    \dot{z} &= \Om p, \\
    \dot{p} &= -\frac{\Oe}{r} z + f_0\left(1-\frac{1}{r^2}\right) + \sqrt{\frac{4\Gqba}{r^2}} X_\inn,\label{eq:p_soft}
    \end{align}
\end{subequations}
where the position and momentum are normalized, respectively, to $\zzpf=\sqrt{\hbar/(m\Om)}$ and $\pzpf=\sqrt{\hbar m \Om}$, $X_\inn(t)$ is the amplitude quadrature of the input field generating photon recoil and  with correlations $\langle X_\inn(t) X_\inn(t')\rangle = \delta(t-t')/2$, $f_0$ is a stray force not originating from the optical tweezer.
We have chosen a reference frame such that its origin corresponds to the equilibrium position of the nanoparticle in absence of a resonance frequency change, i.e., $z_\mathrm{eq}=0$ at $r=1$.
Our experiments are much shorter than the damping time constant due to either gas collisions or radiation damping.
Thus, we safely ignore any damping term.

There are three effects associated with changing the resonance frequency. 
First, the restoring force changes according to the first term on the right-hand-side (RHS) of Eq.~\eqref{eq:p_soft}.
Second, the strength of the photon recoil changes as well, due to the lower optical power impinging on the nanoparticle (last term of Eq.~\eqref{eq:p_soft}).
Third, the stray force shifts the equilibrium position to $z_\mathrm{eq}=f_0/\Om \left(r^2-1\right)$.
This is due to a change in the balance between the stray force ($f_0$, independent of $r$) and the radiation pressure force (proportional to the optical power, thus to $r^{-2}$).

The solution to Eqs.~\eqref{eq: second EOM} is
\begin{subequations}
    \label{eq: general solution}
    \begin{align}
    z(t) &= z_\mathrm{eq}+\left(z_0-z_\mathrm{eq}\right) \cos(\Oe t) + r\,p_0\sin(\Oe t) + \sqrt{4\Gqba} \int_0^t\dd\tau\,\sin\big( \Oe(t-\tau)\big) X_\inn(\tau), \\
    p(t) &= -\frac{z_0-z_\mathrm{eq}}{r} \sin(\Oe t) + p_0\cos(\Oe t) + \sqrt{\frac{4\Gqba}{r^2}} \int_0^t\dd\tau\,\cos\big( \Oe(t-\tau)\big) X_\inn(\tau),
    \end{align}
\end{subequations}
where $z_0$ and $p_0$ are the initial position and momentum values generated by the ``cooling'' step.

Let's first analyze the deterministic part of Eqs.~\eqref{eq: general solution}, i.e., by putting $\Gqba=0$.
These equations describe an elliptical rotation in the $z$-$p$ plane with center at $(z,p)=(z_\mathrm{eq},0)$.
During this rotation, the values of position and momentum interchange each other.
In particular, after a quarter of an oscillation ($t=\pi/(2\Oe)$), the position is an amplified and shifted version of the momentum (i.e., $z_\mathrm{eq}+r p_0$), while the momentum is a de-amplified and mirrored version of the initial position (i.e., $-(z_0-z_\mathrm{eq})/r$).
In essence, this is the basic principle that allows us to delocalize the nanoparticle's wavefunction to lengthscales larger than the zero-point motion.

The stochastic part of Eqs.~\eqref{eq: general solution}, i.e., the terms proportional to $\Gqba$, describe the impact of photon recoil on the state during the evolution.

We compute from Eqs.~\eqref{eq: general solution} the mean values and the covariance matrix for the state as a function of the evolution time.
We assume as initial state a thermal state, with zero mean values $\langle z_0 \rangle=\langle p_0\rangle=0$ and variances $V_z^{(0)}=V_p^{(0)}=\bar{n}+1/2$ and $C_{zp}^{(0)}=0$, where $\bar{n}$ is the average phonon occupancy generated by the "cooling" step and $C_{zp}^{(0)} = \langle \{ \Delta z, \Delta p\}\rangle/2$ is the symmetrized covariance between position and momentum, with $\{\cdot, \cdot\}$ being the anticommutator and $\Delta j = j-\langle j \rangle$ for $j \in\{z,p\}$.
Also, we assume that the stray force $f_0$ is constant during a single repetition of the experiment, but varies between different repetitions. In practice, this translates into fluctuations of the equilibrium position $z_\mathrm{eq}$. We denote by $\langle z_\mathrm{eq}\rangle$ and $V_{eq}$ the ensemble mean and variance, respectively.
The variance of the equilibrium position stems from the fluctuating force, thus $V_\mathrm{eq} = V_{f_0}/\Om^2(r^2-1)^2$.

The mean values of the state are
\begin{subequations}
\label{eq: mean value}
\begin{align}
    \langle z \rangle &= \langle z_\mathrm{eq}\rangle -\langle z_\mathrm{eq}\rangle\cos(\Oe t), \\
    \langle p \rangle &= \frac{z_\mathrm{eq}}{r} \sin(\Oe t).
    \end{align}
\end{subequations}
We use the ensemble average of the position over many realizations to infer the average stray force acting on the particle.
We use this information to apply a static force that cancels the average stray force (see Sec.~\ref{sec: SFC}).
However, fluctuations from one trajectory to another are still present and cannot be compensated with the present method, as it relies on ensemble average.

The covariance matrix $\Sigma$ of the state can be computed from Eqs.~\eqref{eq: general solution} as well.
In our case, this matrix can be written as $\Sigma=\Sigma_\mathrm{coh}+\Sigma_\mathrm{pr} + \Sigma_\mathrm{sf}$, where we separate the contributions stemming from the coherent dynamics (first term) and from the decoherence, induced by the photon recoil (second term) and by the drifts in the stray force (last term).
The elements of $\Sigma_\mathrm{coh}$ are
\begin{subequations}
    \label{eq: sigma coh}
    \begin{align}
    V_{z,\mathrm{coh}} &= V_z^{(0)} \cos^2(\Oe t) + r C_{zp}^{(0)}\sin(2\Oe t) + r^2 V_p^{(0)} \sin^2(\Oe t), \\
V_{p,\mathrm{coh}} &= \frac{V_z^{(0)}}{r^2}\sin^2(\Oe t) - \frac{C_{zp}^{(0)}}{r} \sin(2\Oe t) + V_p^{(0)}\cos^2(\Oe t), \\
C_{zp, \mathrm{coh}} &= \left( -\frac{V_q^{(0)}}{2r} + \frac{r V_p^{(0)}}{2} \right) \sin(2 \Oe t) + C_{zp}^{(0)}\cos(2\Oe t).
    \end{align}
\end{subequations}
From Eqs.~\eqref{eq: sigma coh} we can see clearly the expansion and compression cycles that occur at a rate $2\Oe$. 
Decoherence reduces however the purity of this state, by broadening both position and momentum variances.
The elements of $\Sigma_\mathrm{pr}$ are
\begin{subequations}
\label{eq: sigma dec}
\begin{align}
    V_{z,\mathrm{pr}} &=  \frac{r \Gqba}{\Om}\left( \Oe t - \frac{\sin(2\Oe t)}{2} \right),\\
    V_{p,\mathrm{pr}} &=  \frac{\Gqba}{r \Om}\left( \Oe t + \frac{\sin(2\Oe t)}{2} \right),\\
    C_{zp, \mathrm{pr}} &= \frac{\Gqba}{2 \Om}\left( 1 - \cos(2\Oe t) \right).
\end{align}
\end{subequations}
We can see from Eqs.~\eqref{eq: sigma dec} that the photon recoil induces a monotonous growth in the variances of the position and momentum, thus inducing a steady broadening of the state. This broadening represents the main limit of our expansion protocol on the final purity of the state. 
Finally, the drifts of the stray force are responsible for $\Sigma_\mathrm{sf}$, with elements
\begin{subequations}
\label{eq: sigma f}
    \begin{align}
        V_{z,\mathrm{sf}} &=   \frac{V_{f_0}}{\Om^2}(r^2-1)^2\left( 1-\cos(\Oe t)\right)^2, \\
        V_{p,\mathrm{sf}} &=  \frac{V_{f_0}}{\Om^2}\frac{(r^2-1)^2}{r^2} \sin^2(\Oe t), \\
        C_{zp,\mathrm{sf}} &=   \frac{V_{f_0}}{\Om^2}\frac{(r^2-1)^2}{r} \left(1-\cos(\Oe t)\right)\sin(\Oe t).
    \end{align}
\end{subequations}

\subsection{Limiting the decoherence due to time-varying stray forces: 1-pulse vs 2-pulse protocols}
The simplest approach to delocalize the nanoparticle's wavefunction is to let the particle evolve in the shallower trap for a quarter of the oscillation, $t=\pi/(2\Oe)$, then trap it back in the original potential. We refer to this scheme as a {\it 1-pulse} protocol.
We have implemented this protocol in our experiment. 
We noticed, however, that after this protocol the measured added noise could not be explained by taking into account only photon recoil (cfs. Eqs~\eqref{eq: sigma dec}).
After a systematic study to rule out other possible decoherence mechanisms, our experiments indicated that the source of this excess noise can be ascribed to slow drifts of the stray fields.
Indeed, from Eqs.~\eqref{eq: sigma f} we can extract the covariance matrix elements of the added noise at $t=\pi/(2\Oe)$, that is
\begin{subequations}
\label{eq: sigma f 1pulse with expansion}
    \begin{align}
        V_{z,\mathrm{sf}}^{1-\mathrm{pulse}} &=  \frac{V_{f_0}}{\Om^2} \left( 
 D^2-1 \right)^2, \\
        V_{p,\mathrm{sf}}^{1-\mathrm{pulse}} &=  \frac{V_{f_0}}{\Om^2} \frac{\left( 
 D^2-1 \right)^2}{D^2} , \\
        C_{zp,\mathrm{sf}}^{1-\mathrm{pulse}} &=  \frac{V_{f_0}}{\Om^2} \frac{\left( 
 D^2-1 \right)^2}{D},
    \end{align}
\end{subequations}
where $D=r$ is the delocalization factor, that is the factor by which the position standard deviation has increased at the end of the protocol.
While the position is delocalized at the end of the protocol, the momentum becomes localized as a consequence of the Heisenberg principle.
Thus, this observable is the most susceptible to decoherence.
For large delocalization factors, i.e., $D\gg1$, we notice that the added noise in the momentum variance scales as $D^2$.

To mitigate this source of decoherence, we developed the experimental sequence presented in the main text, that we denote {\it 2-pulse} protocol.
In this sequence, we apply two 1-pulse protocols delayed by a quarter of an oscillation in the original trapping potential (see Fig.~1 of the main text).
The delay changes the sign of the average momentum acquired by the particle during the first pulse due to the stray force, such that during the second pulse the nanoparticle moves closer to the original equilibrium position.
At the same time, the delay interchanges the position and momentum variance, such that the second pulse keeps delocalizing the nanoparticle.

We recursively use Eqs.~\eqref{eq: sigma coh} and~\eqref{eq: sigma dec} to compute the covariance matrix of the coherent part and the decoherence due to photon recoil.
The result is Eqs.~(2) from the main text. 
In particular, we notice that the delocalization factor now scales with $r^2$, that is $D=r^2$.
Similarly, we can compute the effect of the time-varying stray force.
We first use recursively Eqs.~\eqref{eq: general solution}, collect the terms proportional to the force $f_0$, and compute the corresponding ensemble variance.
We find, in terms of the delocalization factor, that 
\begin{subequations}
\label{eq: sigma f 3pulse with expansion}
    \begin{align}
        V_{z,\mathrm{sf}}^{2-\mathrm{pulse}} &= \frac{V_{f_0}}{\Om^2} (D-1)^2(\sqrt{D}-1)^2, \\
        V_{p,\mathrm{sf}}^{2-\mathrm{pulse}} &= \frac{V_{f_0}}{\Om^2} \frac{(D-1)^2(\sqrt{D}-1)^2}{D^2}, \\
        C_{zp, \mathrm{sf}}^{2-\mathrm{pulse}} &= -\frac{V_{f_0}}{\Om^2} \frac{(D-1)^2(\sqrt{D}-1)^2}{D}.
    \end{align}
\end{subequations}
In this case, we notice that the added noise in the momentum variance scales linearly with $D$ for large $D$, and not quadratically as for the 1-pulse protocol.
This means that the 2-pulse protocol provides an advantage in terms of reducing the added noise of technical origin.
In our experiments, at the largest delocalization of $D=6$, we find that the position and momentum terms in Eq.~\eqref{eq: sigma f 3pulse with expansion} are more than twenty times lower than the corresponding terms in Eq.~\eqref{eq: sigma f 1pulse with expansion}, demonstrating the robustness of the 2-pulse protocol against drifting stray forces. 

\clearpage
\newpage
\section{Experimental Details}
In this section we provide more details about the experimental setup and the methods used to prepare the initial state by feedback, to switch the trapping laser power and to compensate the average stray force experienced by the nanoparticle.

\subsection{Experimental setup}\label{sec:expsetup}
\begin{figure}[h]
    \centering
    \includegraphics[width=172mm]{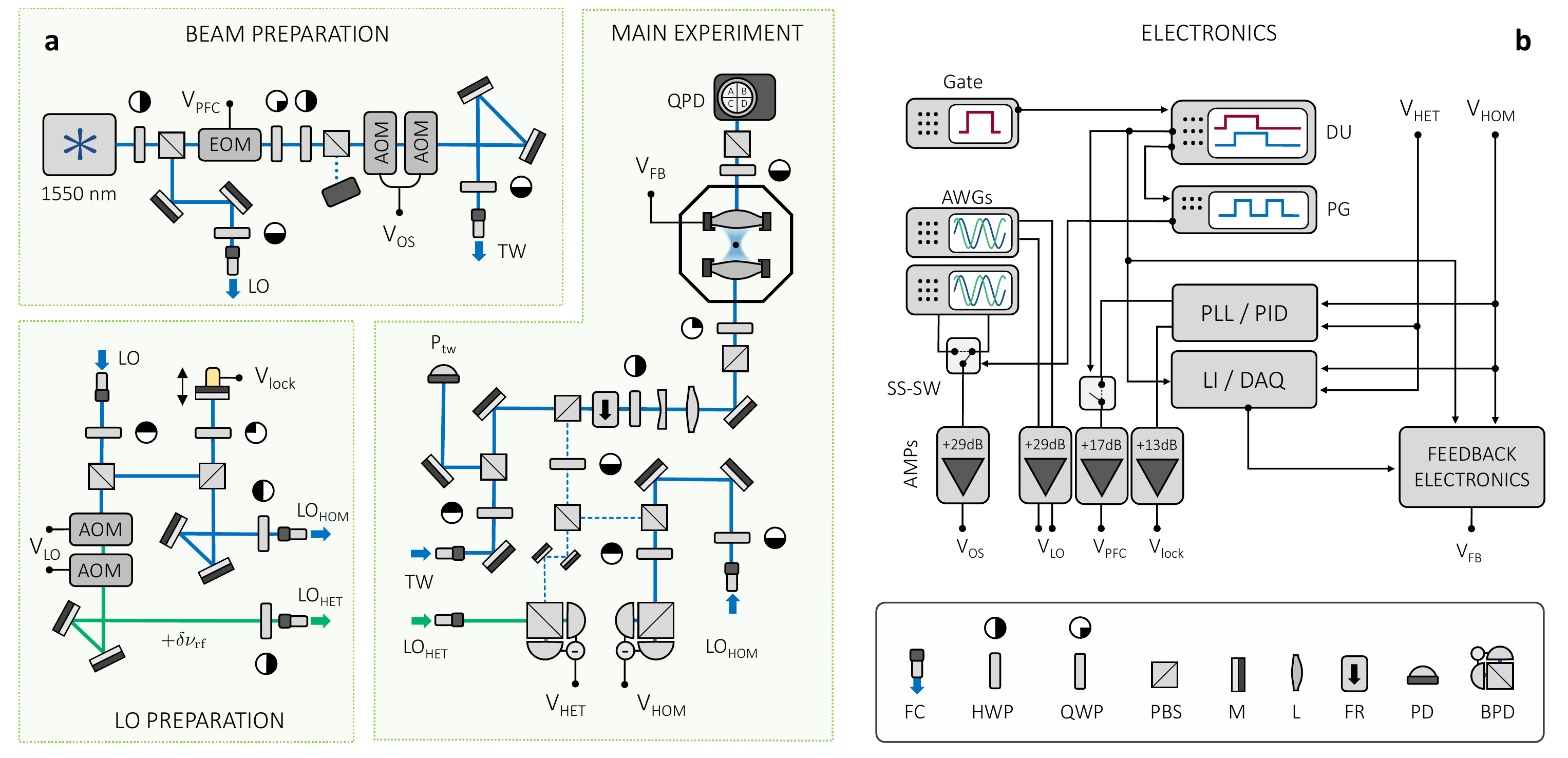}
    \caption{\textbf{Experimental setup overview. a} Optical setup, where we optically trap the dielectric nanoparticle and measure its motion. \textbf{b} Electronic setup, used to cool the particle motion, generate the optical pulse sequence for the expansion protocol, and acquire timetraces.}
    \label{fig: SetupDetails}
\end{figure}

We show the details of the optical setup in Fig.~\ref{fig: SetupDetails}a and of the related electronics in Fig.~\ref{fig: SetupDetails}b. 

The optical setup is based on the one presented in Ref.~\cite{tebbenjohanns_quantum_2021}, and consists of three main sections: the beam preparation, the local oscillator (LO) preparation, and the main experiment. 
In the beam preparation section, we first modulate a \SI{1550}{nm} wavelength laser with an electro-optic modulator (EOM).
Then, we split it in two beams, one used to generate the optical tweezer (TW), and the other used to generate the LOs. 
The tweezer beam goes through an optical switch consisting of two cascaded acousto-optic modulators (AOM), see Sec.~\ref{sec: optical switch}. 
In the LO preparation section, we split the LO beam into two beams. We propagate one beam through two cascaded AOMs in order to detune its frequency: this is the LO for heterodyne detection (LO$_\mathrm{HET}$). The two AOMs are such that the first one upshifts the laser frequency by \SI{80}{MHz} and the second downshifts it by either \SI{79.02}{MHz} or \SI{80.98}{MHz}, depending on the desired detuning of the LO$_\mathrm{HET}$ (see Sec.~\ref{sec: initial phonon occupancy}).
The second beam serves as LO for a homodyne detector. We use a piezo mirror to control and actively stabilize the phase relation between the LO and the optical tweezer.
In the main experiment section, we propagate the tweezer light through an optical circulator (based on a Faraday rotator), and focus it with a 0.8\,NA lens to form an optical tweezer. 
Before the optical circulator, we sample a small fraction of the tweezer beam in order to monitor its power (P$_\mathrm{TW}$). 
We collect the scattered light propagating along the tweezer direction with a quadrant photodiode (QPD) and use it for calibration purposes at high pressures. 
We separate the light scattered against the tweezer direction by collecting it at the third port of the optical circulator (dashed line). 
We split this scattered light into two parts, then overlap one with the LO$_\mathrm{HOM}$ and the other with the LO$_\mathrm{HET}$. 
We then use two balanced photodetectors (BPD) to derive the homodyne (V$_\mathrm{HOM}$) and the heterodyne (V$_\mathrm{HET}$) signals, respectively. 

The electronic setup of Fig.~\ref{fig: SetupDetails}b serves the purpose of controlling the optical tweezer power during the expansion protocol, generating the signal for motional feedback cooling, generating the signal for phase stabilization of the homodyne LO, and acquiring time traces. 
To generate the pulse sequence used in the expansion protocol, we combine a Stanford
Research Systems DG53 pulse delay unit (DU) and an Agilent 81110A pattern generator (PG).
Because the vibrations induced by the cryo-cooler affect both the trap and the detection, it is important to perform the expansion protocol only during the time intervals when such vibrations are minimal. 
To do so, we gate the DU with a square wave at frequency \SI{1}{Hz}, which is the frequency of the compressor cycles.
We adjust the phase and duty cycle of this square wave in such a way that it has high voltage when the vibrations are low and low voltage otherwise. 
We use this gate to allow the generation of pulses for the expansion protocol only when the setup is stable.
We use the gated DU as a master trigger for the rest of the experiment. At each run of the expansion protocol, the DU generates a pulse that switches off the cold damping and parametric feedback cooling signals. 
At the same time, the DU sends a trigger signal to the PG, which outputs a TTL pulse sequence that actuates the electronics responsible for the tweezer power modulation (see Sec.~\ref{sec: optical switch}). 
Finally, the DU generates also a TTL signal that triggers the acquisition of the particle time trace. 

To make use of the pulse signals output by the DU and PG, we make use of Mini-Circuits ZASWA-2-50DRA+ solid state switches (SS-SW). These devices consist of two input RF ports and one RF output port. A further transistor-transistor logic (TTL) port selects which of the input ports is connected with the output. 
To toggle the feedback, we connect the feedback signal to one of the two RF ports, and terminate the second one on a \SI{50}{Ohm} impedance. 
The trigger for the feedback generated by the DU selects then whether to propagate the feedback signal further towards the particle or not. 

In order to toggle the power of the optical tweezer, instead, we follow a slightly different approach. As explained in Sec.~\ref{sec: optical switch}, the power of the trapping beam is controlled by an optical switch whose drive is a \SI{40}{MHz} tone. 
The amplitude of this RF modulation determines the power at the output of the optical switch. Given a desired frequency ratio $r$, we use the calibration described in Sec.~\ref{sec: optical switch} to determine the amplitudes of the RF tone corresponding to the high and to the low mechanical resonances. 
We generate two separate \SI{40}{MHz} tones on two channels of a Keysight 33600A arbitrary wave generators (AWG). We set the amplitude of each channel to be the desired amplitude for the low or high resonance frequencies. 
We input these two tones to the RF ports of the SS-SW. 
The output of the PG that controls the optical power, then, is used as TTL signal that selects between the high and the low RF powers. 
Note that we also align the phases of these two RF signals in order to avoid that the AOMs of the optical switch generate power spikes induced by a discontinuity in the phase of their RF drive tones. 

We use two Zurich Instruments MFLI lock-in devices, each of them receiving the particle signal coming from the homodyne and heterodyne detectors. We use the first MFLI (PLL/PID) to generate the parametric feedback signal V$_\mathrm{PFC}$ (see Sec.~\ref{sec: feedback cooling}) and the lock signal V$_\mathrm{lock}$ for the phase stabilization of the homodyne LO. We use the second MFLI (LI/DAQ) as a data acquisition card triggered by the DU signal. We also use the digital outputs of this second MFLI to control the gain of the linear feedback signal V$_\mathrm{FB}$. 

\subsection{Feedback cooling}
\label{sec: feedback cooling}

As soon as the particle is in ultra-high vacuum, it is necessary to stabilize its motion via feedback cooling in order to minimize the risk of losing it. 
Even though the transversal axes of motion of the particle are not considered in the main text, it is nevertheless crucial to detect and cool them as well. 

We make use of two different cooling techniques. While we are not performing measurements, we perform parametric feedback cooling. 
We achieve this with a phase-locked loop (PLL) implemented on a Zurich Instruments MFLI lock-in box. 
This PLL tracks the motion of the particle along the three axes and produces three corresponding harmonics that are phased locked with the motion. 
We sum these three harmonics on the output of the lock-in box, and feed them into an FPGA (Redpitaya) that produces the square of this signal. 
The resulting signal contains all combinations between the three harmonics of motion. 
Of relevance to us is the fact that the second harmonic of each axis of motion is contained in this output signal, whereas all the other spectral components do not impact the motion in a relevant way. 
We use the output of the FPGA to modulate the power of the tweezer laser via the EOM in the beam preparation section (see Sec.~\ref{sec:expsetup}).
The resulting parametric cooling is relatively mild, and keeps the particle state in a thermal state of a few hundreds of motional quanta. 

The parametric feedback scheme has the advantage of being insensitive to the sign of the motion detection. 
This is important, for instance, when resetting the LO phase locking electronics.
During experiments that require quantum control of the particle motion, such as the delocalization we report in the main text, we cannot rely on this scheme.
In this case, we implement a cooling scheme based on so-called cold damping on all three motional axes. 
We achieve this with a custom-programmed FPGA (Redpitaya) that takes as input the signal derived from the backscattering detection scheme. This signal has a maximal signal to noise ratio for the longitudinal motion, and a high enough measurement quality for the transversal axes to allow for a reliable stabilization. 

\begin{figure}
    \centering
    \includegraphics[width=160mm]{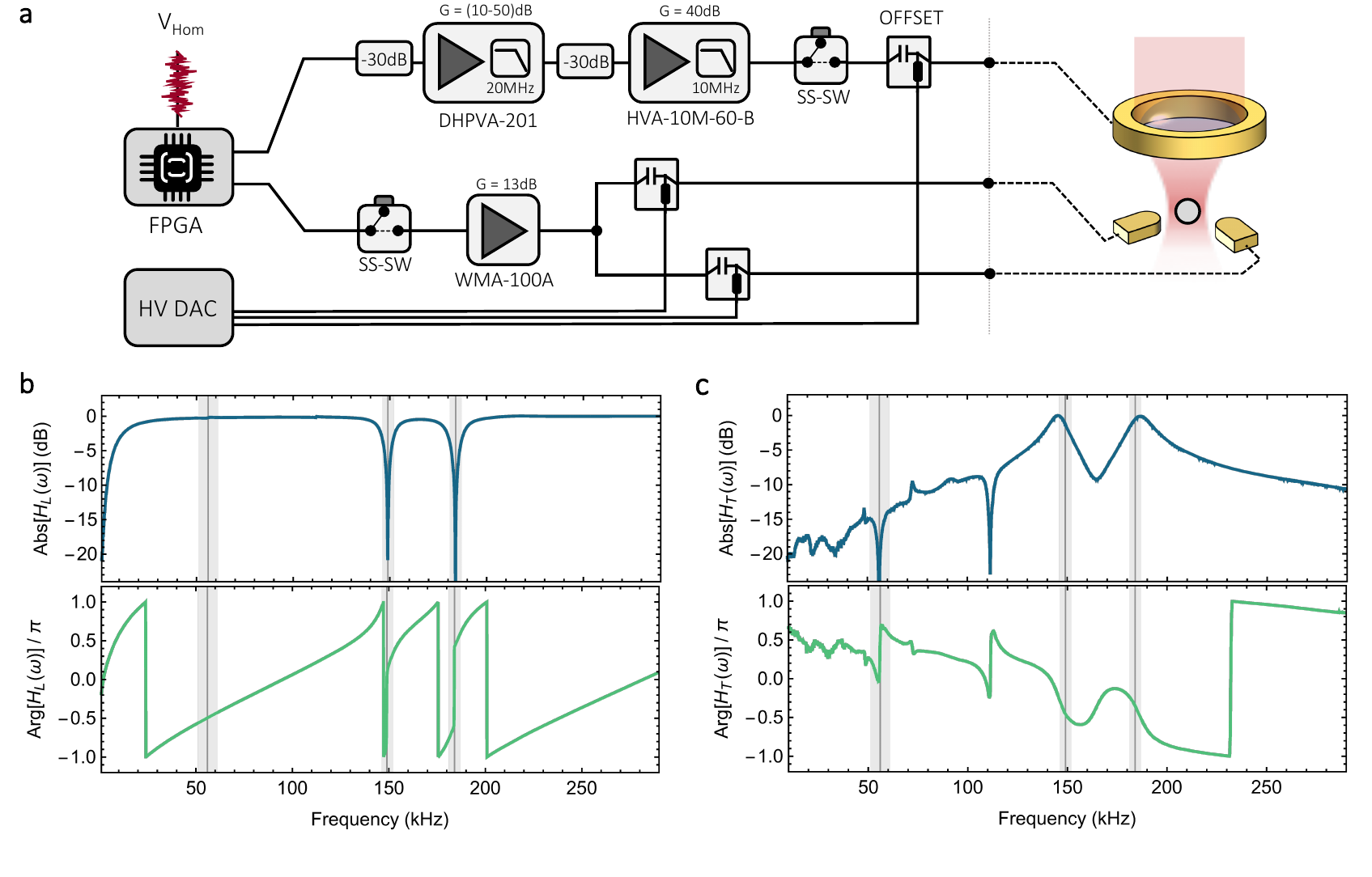}
    \caption{\textbf{Feedback electronics. a} The homodyne position measurement is fed to a custom programmed FPGA  board that produces the feedback signal. This signal is then propagated through an amplification chain. After amplification, we combine this feedback signal with a DC voltage generated with a high voltage digital to analog converter (HV DAC), and apply the output on electrode pairs around the trapped particle. \textbf{b, c} Amplitude and phase of the transfer function used to generate the longitudinal (\textbf{b}) and transverse (\textbf{c}) feedback signals from the position measurement.}
    \label{fig: feedback electronics}
\end{figure}

The cold damping scheme consists of using a combination of linear filters and delays (more details below) in order to derive a signal that mimics a damping force on the particle motion that does not add corresponding fluctuations. 
We amplify this signal with a combination of commercial low noise amplifiers.
Then, we apply it on the particle via three pairs of electrodes, each aligned with one of the motional axes. We show the amplification chain in detail in Fig.~\ref{fig: feedback electronics}a.

In Fig.~\ref{fig: feedback electronics}b and~~\ref{fig: feedback electronics}c, we show the transfer functions of the longitudinal and transversal feedbacks. To derive the longitudinal feedback, we optimize the parameters in order to allow cooling close to the motional ground state. To do so, we apply a second order \SI{10}{kHz} high pass filter in order to eliminate low-frequency electronic noise stemming from the detector, apply two notch filters ($Q$ factor 15) centered at the resonance frequencies of the transversal axes ($\Omega_x = 2\pi\times \SI{149.5}{kHz}$ and $\Omega_y = 2\pi\times \SI{183.7}{kHz}$), and a delay of $\SI{4.78}{\mu \mathrm{s}}$. The notch filters are required in order to avoid feeding back information related to the transversal motion, since the nonzero components of the electric field along these axes may induce undesired heating. The role of the electronic delay, instead, is to make sure that the phase of the transfer function at the longitudinal resonance frequency is \SI{90}{deg}, allowing us thus to approximate a derivative filter that damps the motion. 

To derive the transversal feedback, instead, we use a first order \SI{10}{kHz} high pass filter, a notch filter ($Q$ factor 10) at \SI{56}{kHz} and \SI{112}{kHz}, corresponding to the first and second harmonics of the longitudinal motion, and a bandpass filter ($Q$ factor 16) centered at \SI{144.5}{kHz} and \SI{183.7}{kHz} for the feedback along the $x$ and $y$ axes, respectively. We notch away both first and second harmonics of the longitudinal motion because the input signal is optimized for this degree of freedom, and both harmonics become dominant if the particle heats up sufficiently. The role of the bandpass filters, instead, is to tune the gain of the filter and provide a phase shift that, combined with the intrinsic delay of the signal propagation, approximates in a narrow band a derivative filter.

\subsection{Optical switch}
\label{sec: optical switch}

We implement the optical switch by cascading two acousto-optic modulators (AOMs), model IntraAction Corp. ACM-402AA1, which are driven by the same RF tone at \SI{40}{MHz}. 
We generate the RF tones with a Keysight 33600A function generator whose output is amplified by a  RF amplifier Mini-Circuits ZHL-3A-S+.
The AOMs are aligned such that the diffracted beam is first up-shifted by the RF drive frequency and then down-shifted by the same amount.
Therefore, the beam at the output of the two AOMs has the same initial frequency, and a power that depends on the amplitude of the RF tone driving the modulators. 
The rise and fall times of the AOMs is $\sim$\SI{250}{ns}, which is much faster than the characteristic timescales of the particle.
Using the same RF tone to drive both AOMs also allows us to suppress common mode phase noise due to the RF source, which is canceled in diffracted beams of the same order but opposite sign.

By amplitude modulation of the RF mode with a square wave, we generate the pulse sequence used in the expansion protocol. 
It is crucial that the two AOMs act on the optical beam at the same time, otherwise the power of the output beam changes from one value to another by first passing through an intermediate value. This intermediate value corresponds to one of the AOMs acting on the light before the other.
We ensure the simultaneous action of the two AOMs by making sure that the acoustic wave generated inside each crystal travels the same distance $d$ before reaching the light beam. Since the speed of sound inside the active crystal is roughly 2.5\,mm/$\mu$s, for instance, we need to adjust $d$ with a precision better than 625\,$\mu$m in order to make sure that the two AOMs act within \SI{250}{ns} from each other (corresponding to their rise time). 

\begin{figure}
    \centering
\includegraphics[width = 3.2 in]{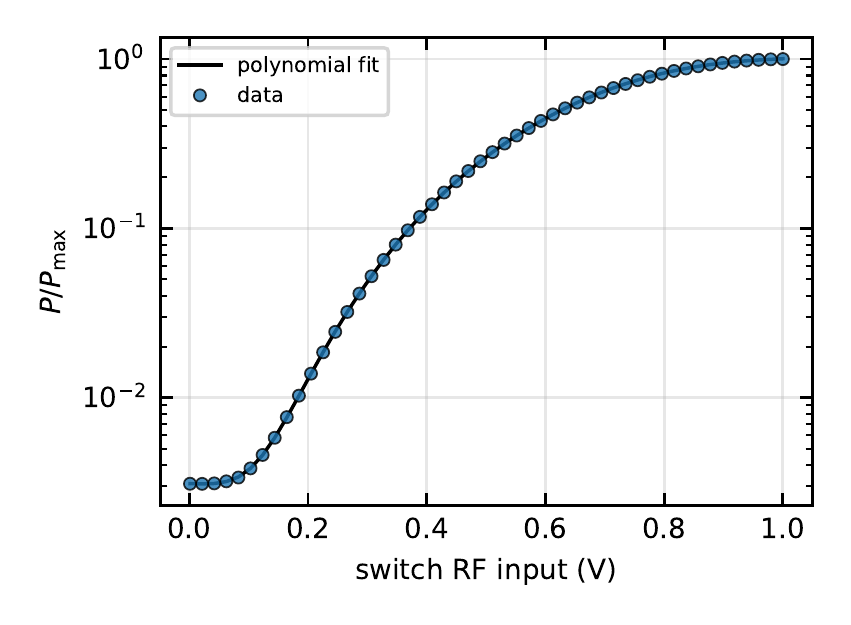}
    \caption{\textbf{Optical switch calibration.} Power of the optical tweezer as a function of the amplitude of the RF tone fed to the optical switch.}
    \label{fig: aom_switch_calibration}
\end{figure}

To calibrate the output optical power against the amplitude of the input RF tone, we direct a portion of this light towards a free space detector. We then apply short pulses---on the order of tens of microseconds---on the RF tone, and record the output power as a function of input voltage. 
We then fit the resulting trend with a seventh order polynomial.
We show an example of such fit in Fig.~\ref{fig: aom_switch_calibration}.
The square root of the optical power, in turn, is proportional to the mechanical resonance frequency of the particle. 
In the experiment, we invert numerically the seventh order polynomial to determine the voltage required to achieve a desired ratio $r$ between the initial and final resonance frequencies. 
For our calibration, it is important to change the amplitude of the RF tone with fast pulses and record the time dependence of the optical power. 

\subsection{Compensation of stray forces}\label{sec: SFC}

In Sec.~\ref{sec:theory_eom} we have seen how a stray force shifts the equilibrium position of our nanoparticle.
This shift is rather small when we trap with the maximum tweezer power, but becomes larger and larger for increasing values of $r$, i.e., for shallower traps.
In practice, we found that the shift of the equilibrium position can be so large that the particle falls outside of the trapping volume, leading to a particle loss.
Hence, it is necessary to counteract these stray force along the three directions.
In this section, we describe how we accomplish this by using static electric fields. 
We describe the procedure we use for the longitudinal motion.
For the transverse axes, the procedure is identical.

\begin{figure}
    \centering
    \includegraphics[width=3.5 in]{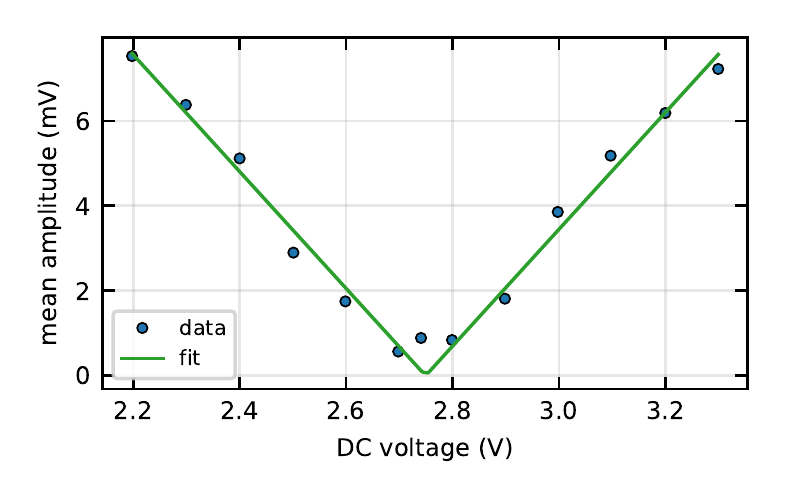}
    \caption{\textbf{Stray field compensation.} Mean oscillation amplitude after frequency jump protocol as a function of an externally applied DC voltage. We observe that there is a clear minimum of the amplitudes, which corresponds to a minimization of the stray force acting on the particle.}
    \label{fig: SFC}
\end{figure}

We first initialize the particle state to a low phonon occupancy through feedback cooling. Then, at time $t=0$, we switch off the feedback and set the power of the optical tweezer to lower values, inducing a new resonance frequency $\Oe = \Om/r$. 
We typically start from $r=1.5$, in order to minimize the risk of particle loss. 
We then let the state evolve under this condition for a half period time, $\Oe t = \pi$, and set then again the power of the tweezer to the original value. 
We let the nanoparticle move in absence of feedback for $\sim1\,\mathrm{ms}$ and we measure its displacement.
We collect an ensemble of 20 such time traces, which we found to be sufficient for the purpose.
We then implement, in post-processing, an anti-causal second order Butterworth bandpass filter of \SI{2}{kHz} bandwidth around the resonance frequency $\Om$ and apply it on each trajectory.
We use the filtered traces in order to extract the amplitude of the harmonic oscillations after we set the power back to the original value. 
Then, we perform an ensemble average of the oscillation amplitude. 
From Eqs.~\eqref{eq: mean value}, assuming $\langle z_0 \rangle = \langle p_0 \rangle = 0$ and setting $\Oe t = \pi$, we expect the mean oscillation amplitude $A$ to be
\begin{equation}
\label{eq: mean amplitude}
A = \frac{2r^2}{\Om} \vert \langle f \rangle \vert,
\end{equation}
with $f=f_0(1-r^{-2})$ the stray force acting on the particle. Equation~\eqref{eq: mean amplitude} shows that the mean amplitude is linearly proportional to the absolute value of the stray force $f$. 
We then apply an external DC voltage on the electrodes that surround the particle, and measure again the mean oscillation amplitude as a function of the voltage. 
This DC voltage is generated by a custom made high-voltage source. 
By minimizing this residual amplitude, we ensure that we have compensated for the (average) stray force acting on the particle. 
We can then increase the sensitivity of our estimation by repeating the compensation protocol for higher values of $r$. We show the value of $A$ as a function of the applied DC voltage for $r=2$ in Fig.~\ref{fig: SFC}, where we also show a fit based on Eq.~\ref{eq: mean amplitude}.
In general, we found that such compensation requires additional fine correction over a timescale of few tens of minutes. 

Note that this stray field compensation voltage can be occasionally high (several tens of volts). Because we apply this compensation voltage on the same electrodes that we use to perform feedback cooling, it is important to decouple the FPGA that generates the feedback from the high voltage source. 
To do so, we use a capacitance in series with the feedback signal. The value of this capacitance (50 nF) is selected such that the higher frequencies of the feedback signal propagate seamlessly through it, while the DC line, in series with a $20\,\mathrm{k\Omega}$ resistor, is blocked. 

\clearpage
\newpage
\section{Data analysis}
In this section we present the methods used to analyze the data.
We show how we extract the initial phonon occupation after feedback cooling, how we filter and extract the best estimate for the nanoparticle position and how we subtract the estimation noise from the data.

\subsection{Initial phonon occupancy }
\label{sec: initial phonon occupancy}
We measure the average occupation of the nanoparticle motion (along the longitudinal axis) by using heterodyne thermometry techniques, which have been detailed in Ref.~\cite{tebbenjohanns_quantum_2021}.

\begin{figure}[h]
    \centering
    \includegraphics[width=172mm]{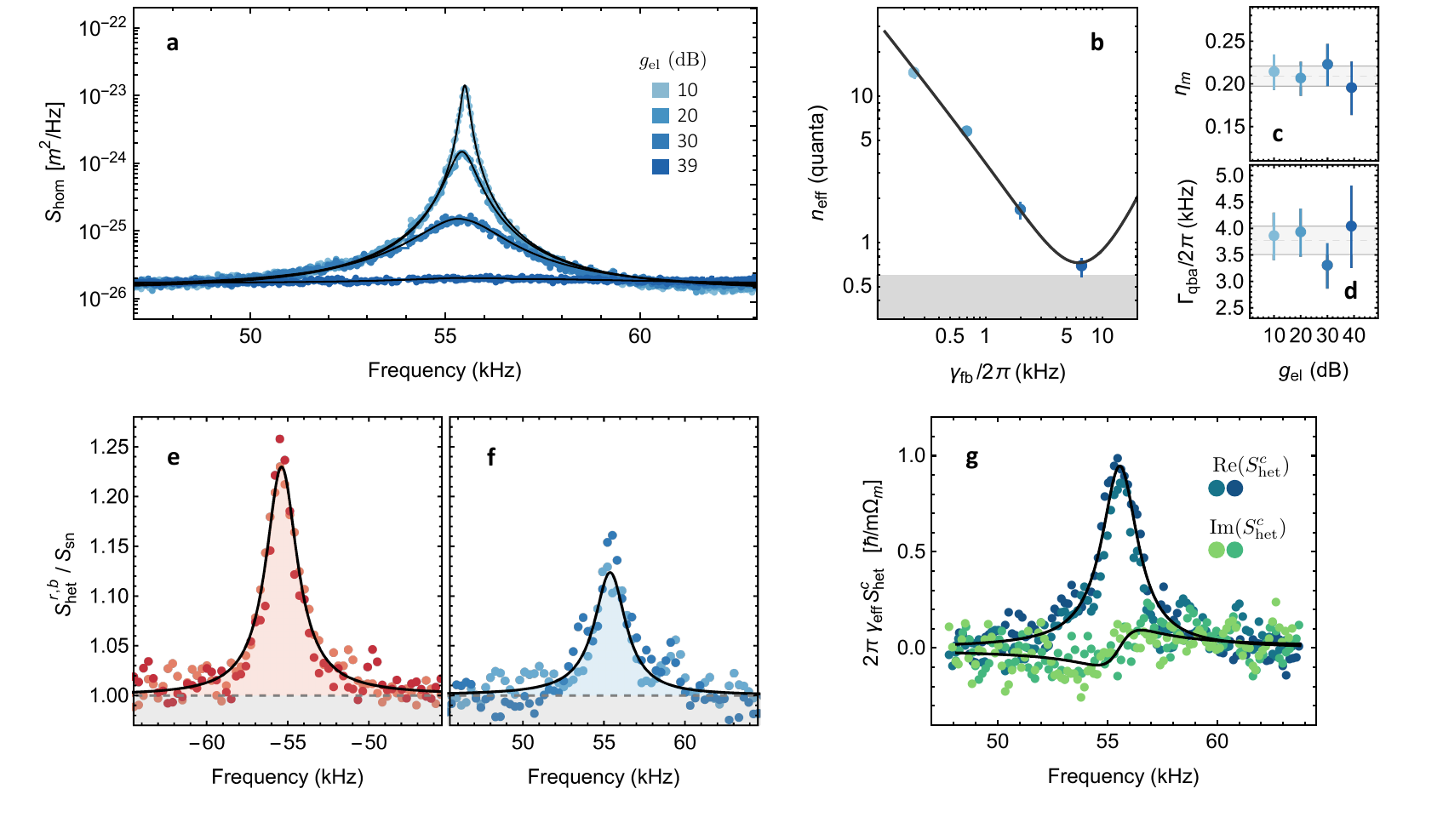}
    \caption{\textbf{Feedback cooling to the ground state.}
    {\textbf a} In-loop displacement spectra at different feedback strengths.
    {\textbf b} Average phonon occupancy.
    {\textbf c, d} Measurement efficiencies and total decoherence rates, extracted from model fitting.
    {\textbf e, f} Red and blue heterodyne sidebands for the largest feedback strength.
    {\textbf g} Spectrum of the red-blue sidebands correlations. Blue (green) circles come from the real (imaginary) part of the cross power spectral density.
    In {\textbf e}, {\textbf f} and {\textbf g} the two shades of colored circles represent two different datasets, acquired with the heterodyne LO frequency above and below the tweezer frequency.}
    \label{fig: ColdDamping}
\end{figure}

In Fig.~\ref{fig: ColdDamping}a, we show the in-loop displacement spectra from a series of feedback cooling experiments, for different feedback strengths (here quantified in $\mathrm{dB}$).
For each experiment, we also measure simultaneously a heterodyne spectrum.
From it, we can extract the red and blue sidebands (Fig.~\ref{fig: ColdDamping}e and f) and their cross correlation (Fig.~\ref{fig: ColdDamping}g).
From fitting the in-loop homodyne spectrum and the heterodyne spectra, we extract the average phonon occupations and fit the four values together.
The results, for each feedback strength, are shown in Fig.~\ref{fig: ColdDamping}b and reported in the table below.
The error bars define the 1-sigma confidence interval from the fits.
\begin{center}
    \begin{tabular}{c|cccc}
    \hline
     $g_\text{el}~\text{(dB)}$  &  10 & 20 & 30 & 39\\
    \hline     
    $\overline{n}_\text{eff}$ & $14.2\pm 0.8$ & $5.65 \pm 0.35$ & $1.66\pm 0.20$ & $0.68 \pm 0.09$\\
    \hline     
    \end{tabular}
\end{center}

From the fits, we also extract the measurement efficiency $\eta_m$ (Fig.~\ref{fig: ColdDamping}c) and the total decoherence rate $\Gtot$ (Fig.~\ref{fig: ColdDamping}d).
The weighted averages are, respectively, $\eta_m=0.21 \pm 0.01$  and $\Gtot/2\pi = \left(3.77 \pm 0.26\right)\,\mathrm{kHz}$; the errors define a 1-sigma confidence interval.

\subsection{Estimation of trajectories}
The measured signal, $i$, linearly depends on the nanoparticle's position and is also affected by noise.
We analyze the measured signal offline to extract the nanoparticle trajectory during the ``measurement" step.
There are two important sources of noise we have to deal with: technical noise coming from electronics and interferometer lock, and shot noise.
The first source of noise affects the low-frequency part of the spectrum, below $5~\mathrm{kHz}$.
Since this frequency band  is well separated from the band of interest for the  motion, centered around $56~\mathrm{kHz}$, we can apply a filter to suppress the technical noise.
We concatenate two second-order high-pass filters, with cutoff frequency of $3~\mathrm{kHz}$ and $9~\mathrm{kHz}$ \cite{rossi_observing_2019}.
We apply these filters offline and backwards in time.

The second source of noise is laser shot noise, which is present for all frequencies (white noise).
We apply a retrodiction filter to extract the best estimate of position at the beginning of the ``measurement" step \cite{lammers_quantum_2024}.
For our case, this filter is essentially a Kalman filter running backwards in time.
In the derivation of the following equations, we assume that the damping rate is negligible and the only decoherence comes from photon recoil.
We define as $\tilde{z}$ and $\tilde{p}$ the best retrodicted estimates of position and momentum, respectively.
These estimates are defined according to
\begin{subequations}\label{eq:conditional_means_retro}
    \begin{align}
        -\dd\tilde{z}&=-\Om\tilde{p}\dd t-8\Gmeas \widetilde{V}_z\tilde{z}\dd t+\sqrt{8\Gmeas}\widetilde{V}_zi \dd t,\\
        -\dd\tilde{p}&=\Om\tilde{z}\dd t-8\Gmeas\widetilde{C}_{zp}\tilde{z}\dd t+\sqrt{8\Gmeas}\widetilde{C}_{zp}i\dd t,
    \end{align}
\end{subequations}
where $-\dd\tilde{z}(t)=\tilde{z}(t-\dd t)-\tilde{z}(t)$ and similarly for $\dd \tilde{p}$.
The measurement outcomes are $i\,\dd t=\sqrt{8\Gmeas}\tilde{z} + \dd W$, where $\dd W$ represent time increments of a Wiener process.
These equations can be interpreted as filter equations that can be evaluated for each realization.
The elements $\widetilde{V}_z$,$\widetilde{V}_p$ and $\widetilde{C}_{zp}$ form the conditional covariance matrix, which evolves over time according to a Riccati equation \cite{lammers_quantum_2024}.
They can be interpreted as the systematic error introduced from the filter, due to residual shot noise that cannot be filtered out.
Our retrodiction lasts for $1\,\mathrm{ms}$, much longer than the timescale associated to Eqs.~\eqref{eq:conditional_means_retro}.
Then, we solve the equation of motion of the conditional covariance matrix at steady-state. We find that the elements of this steady-state matrix are
\begin{subequations}\label{eq:cond_cov}
    \begin{align}
        \widetilde{V}_z&=\frac{\sqrt{-1+\sqrt{1+\eta\Lambda^2}}}{\sqrt{2}\eta\Lambda},\\        \widetilde{V}_p&=\sqrt{1+\eta\Lambda^2}\frac{\sqrt{-1+\sqrt{1+\eta\Lambda^2}}}{\sqrt{2}\eta\Lambda},\\
        \widetilde{C}_{zp}&=-\frac{-1+\sqrt{1+\eta\Lambda^2}}{2\eta\Lambda}
    \end{align}
\end{subequations}
where $\Lambda=4\Gqba/\Om$.
From $\eta$ and $\Gtot$ extracted from the feedback cooling data (see Sec.~\ref{sec: initial phonon occupancy}), we compute Eqs.~\eqref{eq:cond_cov} and find that $\widetilde{V}_z\approx\widetilde{V}_p=1.1$ and $\widetilde{C}_{zp}=0.03$.
In the main text, we have referred to the diagonal elements ($\widetilde{V}_z$ and $\widetilde{V}_p$) as the estimation noise, $V_n$.

For the sake of computational speed, we apply the filter in Eqs.~\eqref{eq:conditional_means_retro} as an infinite impulse response (IIR) filter using the Python scipy module \cite{virtanen_scipy_2020} starting from the most recently acquired data point and working backwards.
We derive the polynomial transfer function representation from the filter's poles and the zeros.
To find them, we apply the Fourier transform to Eqs.~\eqref{eq:conditional_means_retro} and solve for $\tilde{z}(\Omega)$ and $\tilde{p}(\Omega)$ as a function of $i(\Omega)$, where $\Omega$ is the angular frequency.
We indicate the Fourier transforms with the argument $\Omega$.
The transfer function has two poles and one zero, $\omega_\pm^{(p)}$ and $\omega^{(z)}$ respectively.
For the position estimation filter, we have
\begin{subequations}
\begin{align}
    \omega_{\pm}^{(p)} &= \imath 4\Gmeas\widetilde{V}_z \pm \sqrt{\Om^2-8\Gmeas\widetilde{C}_{zp}\Om - 16\Gmeas^2\widetilde{V}_z^2},\\
    \omega_z^{(z)}&=-\imath\frac{\widetilde{C}_{zp}}{\widetilde{V}_z}\Om,
\end{align}
\end{subequations}
where we used $\imath=\sqrt{-1}$.
The momentum filter has the same poles but a different zero, that is
\begin{eqnarray}
    \omega_p^{(z)}&=\imath\frac{\widetilde{V}_z}{\widetilde{C}_{zp}}\Om.
\end{eqnarray}

\subsection{Fit and calibration}
\begin{figure}[t]
    \centering
    \includegraphics[width=172mm]{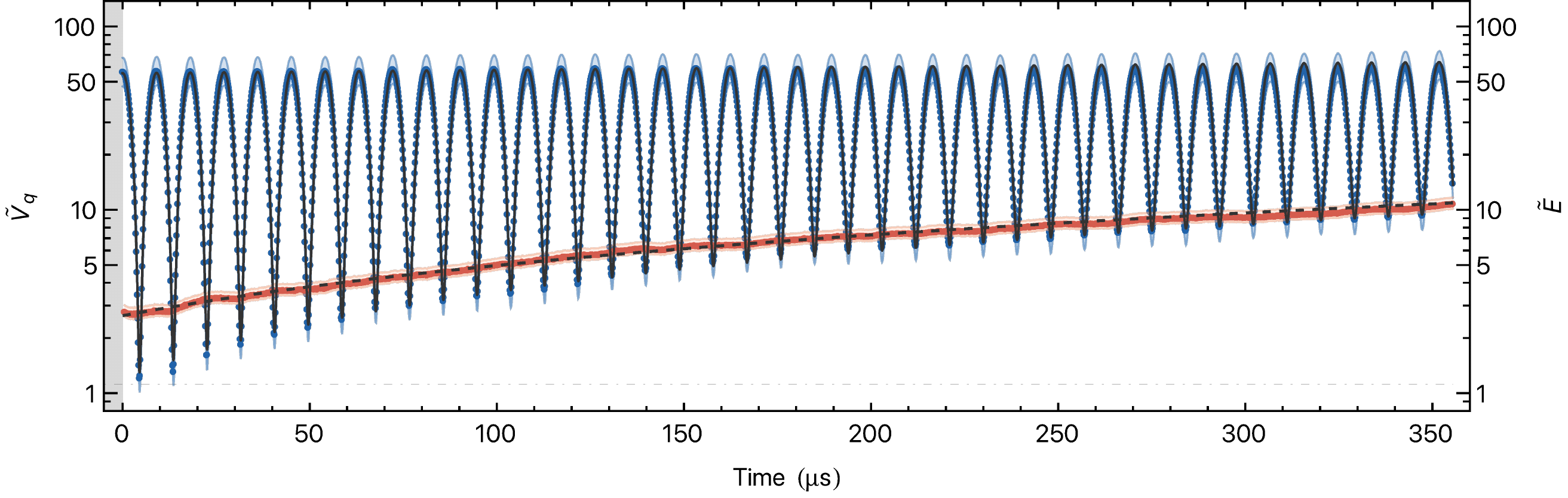}
    \caption{\textbf{Fit of the estimated position variance and energy}:  
    Time evolution of the position ensemble variance ($\tilde{V}_q$, blue circles), and energy after recompression ($\tilde{E}$, tangerine circles). 
    Shaded areas correspond to the 95\% CL intervals of the ensemble variance estimate, the solid (dashed black) line represents the best fit to the position variance (energy) data.
    The dot-dashed line represents the bias noise introduced by the filter. 
    Time $t=0$ correspond to the end of both the expansion and recompression protocols.
    Here the oscillator is initially prepared with an average occupation of $\overline{n}=0.7\pm0.1$ ($g_\text{el}=39~\mathrm{dB}$) and $r^4=36$.}
    \label{fig: fitVE}
\end{figure}
Examples of filtered trajectories are shown in Fig.~1a of the main text.
To convert these data from a voltage into units of $\zzpf$, we use the following procedure.
First, we calculate an ensemble variance for the trajectories, as shown in Fig.~1b of the main text.
Then, we fit these data (from $0$ to $360~\mathrm{\mu s}$) to Eq.~(1) of the main text. 
The fitting parameters are the three elements of the covariance matrix, $V^{\mathrm{volt}}_z$, $V^{\mathrm{volt}}_p$, $C^{\mathrm{volt}}_{zp}$, and the decoherence rate, $\Gamma^{\mathrm{volt}}_\mathrm{tot}$.
The first three parameters come from the maximum and minimum values and the initial phase of the oscillations shown in Fig.~1b of the main text.
The last parameter is the slope with which the variance minima increase over time.
These fitted parameters are, at this point, in units of $V^2$ and $V^2/s$ respectively.
To calibrate the data in units of $\zzpf$, we now assume that the observed decoherence is identical to the decoherence measured in  our ground-state cooling experiments..
That is, we use the factor $\Gtot/\Gamma^{\mathrm{volt}}_\mathrm{tot}$ as our calibration factor to convert the variances $V^{\mathrm{volt}}_z$, $V^{\mathrm{volt}}_p$, $C^{\mathrm{volt}}_{zp}$ into $\widetilde{V}_z$, $\widetilde{V}_p$, $\widetilde{C}_{zp}$.
These three numbers form the estimated covariance matrix of the delocalized state.
Due to inaccuracies of the pulse durations and to the decoherence during the pulse sequence, the covariance matrix is not exactly diagonal.
We can depict this matrix as an ellipse in a $z$-$p$ space.
Being not diagonal corresponds to having the ellipse with the major axes not horizontal.
One can render it diagonal by letting the nanoparticle evolve in a harmonic potential for a fixed amount of time.
To extract the maximum achievable delocalization, we then compute the eigenvalues of the estimated matrix. 
We can also derive the tilt angle of the covariance matrix.
We find an average tilt angle $\overline{\theta}_\text{tilt}=(0.015\pm0.005)\pi$.

To show that the energy is preserved during our protocol,
we perform a different experiment in which the state is delocalized and compressed back (see main text for more details).
To estimate the energy (gray points in Fig.~3 of the main text), we also make use of the momentum estimation.
The procedure is similar to before: from each realization, we apply now two filters (one for position and one for momentum).
We calculate the ensemble variances and we average them, to extract $\widetilde{E} = (\widetilde{V}_z + \widetilde{V}_p)/2$.
To calibrate the data, we use the same calibration coefficient found in the delocalization experiment.
This is justified because we perform the loop experiment right after the delocalization one.
We fit the initial $360\,\mathrm{\mu s}$ of the calibrated estimated energy to
\begin{eqnarray}
    \widetilde{E}(t)=\widetilde{E}_0+\Gqba t
\end{eqnarray}
where the only fitting parameter is the initial estimated energy, $\widetilde{E}_0$. 
The fitted parameter, for each $r$, is what we show as gray points in Fig.~3 of the main text.

\subsection{Subtraction of bias noise introduced by the filter}
\label{sec: noise subtraction}

Our position readout adds a minimum amount of noise to both position and momentum estimation.
This noise is represented by the conditional variances in Eqs.~\eqref{eq:cond_cov} and amounts to additional $V_n=1.1$ phonons to both position and momentum measurements.
Nevertheless, if one calibrates the value of $V_n$, then it is possible to subtract it from the estimated covariance matrix and retrieve the one unbiased by the readout noise.

\begin{figure}
    \centering
    \includegraphics[width=160mm]{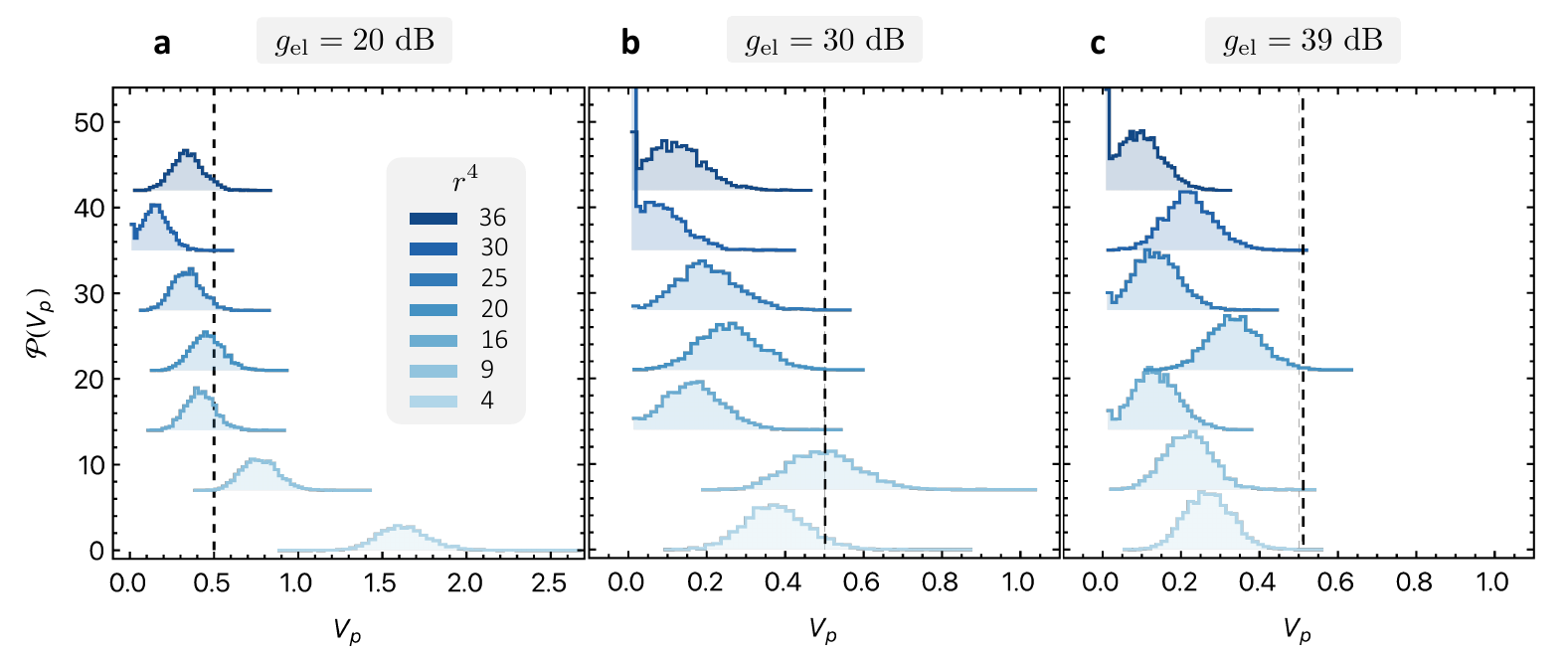}
    \caption{\textbf{Raw Montecarlo histogram of reconstructed momentum variance.} The panels \text{a, b, c} refer, respectively, to feedback gains of $g_\mathrm{el}=\SI{20}{dB}$, $g_\mathrm{el}=\SI{30}{dB}$, and $g_\mathrm{el}=\SI{39}{dB}$. For each subpanel, we show the reconstructed Montecarlo histograms for each value of the frequency ratio $r$. The vertical offset between different distributions is added manually for enhanced readability. The vertical black dashed line represents the momentum variance of the motional ground state.}
    \label{fig: rawMChistograms}
\end{figure}

In order to obtain, after subtraction, an estimation of a physical covariance matrix, i.e. positive semidefinite and satisfying the Heisenberg uncertainty principle, we implement a positive semidefinite optimization, as outlined for instance in Ref.~\cite{kotler_direct_2021}.
The optimization tries to find the covariance matrix $\mathcal{C}$ that minimizes $||\mathcal{C}_\mathrm{meas}-(\mathcal{C}+\widetilde{\mathcal{C}}) ||_2$ with the constraint that $\mathcal{C}+\imath \Omega/2\ge0$, where $\mathcal{C}_\mathrm{meas}$ is the estimated covariance matrix, $\widetilde{\mathcal{C}}$ is the conditional covariance matrix with elements defined in Eqs.~\eqref{eq:cond_cov} and $\Omega$ is the symplectic matrix. 
The notation $||\cdot||_2$ indicates the $l_2-$norm.
To estimate the uncertainty of the covariance matrix after noise subtraction, we perform a Montecarlo estimation. 
To do so, based on $\mathcal{C}_\mathrm{meas}$, we generate 3200 realizations of a multivariate Gaussian distribution that has the same mean and variance as our estimated covariance matrix.
For each of these realizations, we perform the reconstruction outlined above of the physical covariance matrix. 
Then, we reconstruct the probability distribution of the resulting values through a histrogram.
From this histogram, we can extract the average value and the 95\% confidence interval. 
The probability distributions of the momentum variance after noise subtraction are shown in Fig.~\ref{fig: rawMChistograms} for three different initial phonon occupations (see Sec.~\ref{sec: extended dataset}).
%

\section{Extended dataset}
\label{sec: extended dataset}

In this section, we analyse the expansion protocol when applied to initial states with different phonon occupancies. 
We tune this initial occupancy by changing the gain of the electronic feedback cooling applied to the particle. 
Specifically, in the main text we use a gain $g_\mathrm{el} = \SI{39}{dB}$ with respect to a fixed reference value. 
This gain initializes the motional state to an occupancy of $\overline{n} = 0.68\pm0.09$. We now consider the cases of $g_\mathrm{el} = \SI{20}{dB}$ and $g_\mathrm{el} = \SI{30}{dB}$, which initialize the particle state to occupancies of $\overline{n} = 5.65\pm0.35$ and $\overline{n} = 1.66\pm0.20$, respectively.
\begin{figure}[ht!]
    \centering
    \includegraphics[width=172mm]{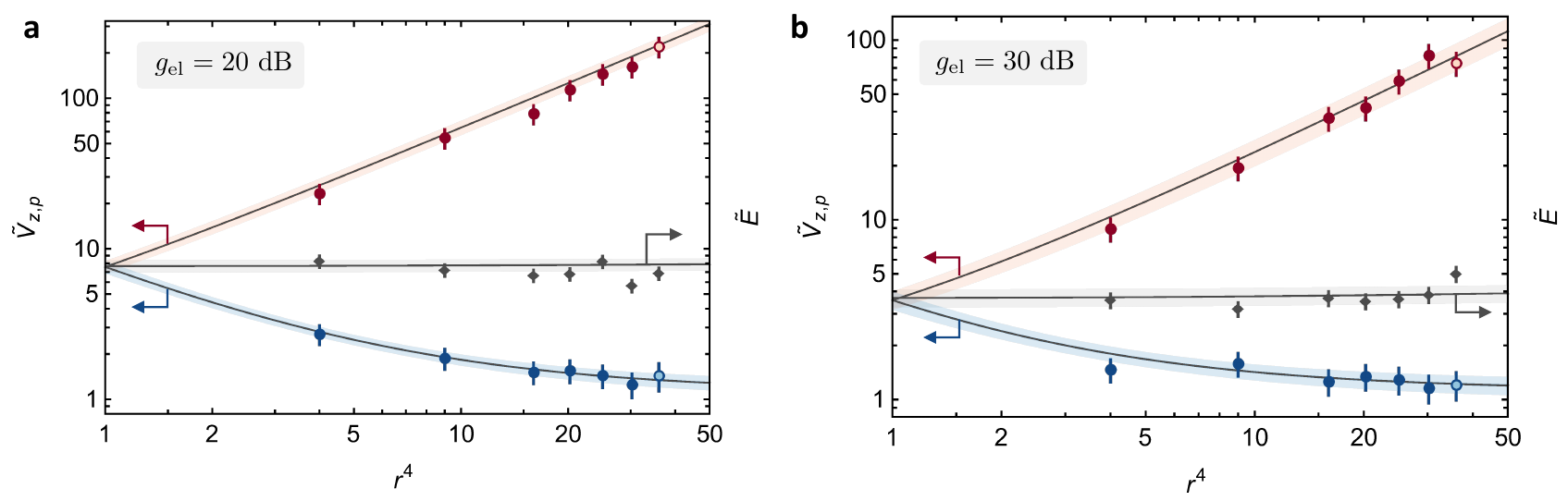}
    \caption{\textbf{Wavefunction expansion and recompression: extended dataset.} Red circles represent the expanded position variance, blue circles the squeezed momentum variance (including measurement noise), solid lines represent the theoretical expectation, while error bars and shaded areas represent the 95\% confidence interval on experimental estimations and theoretical predictions. \textbf{a} Dataset with electronic gain $g_\mathrm{el}=\SI{20}{dB}$.
    \textbf{b} Dataset with electronic gain $g_\mathrm{el}=\SI{30}{dB}$.}
    \label{fig: extended Fig 3}
\end{figure}%
Apart for the initial state, the rest of the experiment and analysis is unaltered from what is presented in the main text. 
For each value of $r$, we repeat the experiment 400 times, apply an optimal retrodiction filter to each trajectory, perform an ensemble variance and use it to extract the covariance matrix of the expanded state. 

For each of the gain values, we show in Fig.~\ref{fig: extended Fig 3} the position and momentum variances after the expansion protocol (red and blue circles, respectively), and the energy after the recompression protocol (grey diamonds). We also overlay the experimental estimations with theoretical curves based on Eqs.~(2) and~(3) from the main text.
Similar to the case of $g_\mathrm{el}=\SI{39}{dB}$, we observe good agreement between theory and experiment. Specifically, the position variance grows with $r^4$ after the expansion protocol, while the momentum variance approaches asymptotically the estimation noise $V_n$. The energy after recompression, in contrast remains the same as the one of the initial state within experimental uncertainty. 

\begin{figure}
    \centering
    \includegraphics[width=172mm]{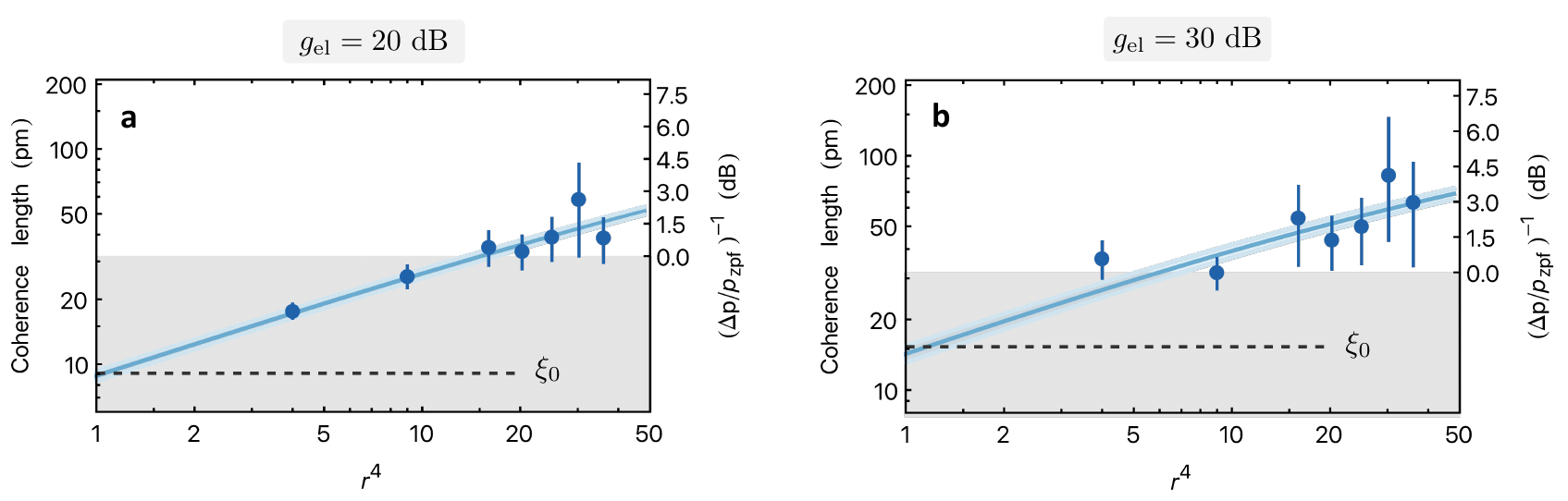}
   \caption{\textbf{Inferred coherence length: extended dataset.} From the estimated momentum variance we subtract the estimation noise and infer the coherence length. The error bars represent the 95\% confidence interval, obtained via a Montecarlo method outlined in Sec.~\ref{sec: noise subtraction}. The light blue solid line is the prediction from our model, and the shaded area reflects the 95\% CL in the model parameters. The shaded gray area bounds the region of coherence lengths that can be obtained without squeezing, while the black dashed line shows the initial coherence length. \textbf{a} Dataset with electronic gain $g_\mathrm{el}=\SI{20}{dB}$.
    \textbf{b} Dataset with electronic gain $g_\mathrm{el}=\SI{30}{dB}$.}.
    \label{fig: extended Fig 4}
\end{figure}
We now extract the true covariance matrix of the expanded states using the same Montecarlo approach discussed in Sec.~\ref{sec: noise subtraction}.
From these matrices, we infer the coherence lengths, which we show in Fig.~\ref{fig: extended Fig 4}.

\bibliography{supplemental.bib}
\bibliographystyle{naturemag}